\documentclass[pra,twocolumn,preprintnumbers,amsmath,amssymb,nofootinbib,floatfix]{revtex4}

\usepackage{graphicx,bm}

\makeatletter
\def\graphicscale{\twocolumn@sw{0.3}{0.4}}
\def\graphicthreescale{\twocolumn@sw{0.3}{0.4}}

\begin{document}

\title{Particle-number scaling of the quantum work statistics and Loschmidt
  echo \\

in Fermi gases with time-dependent traps}

\author{Ettore Vicari}

\affiliation{Dipartimento di Fisica dell'Universit\`a di Pisa and
  INFN, Largo Pontecorvo 3, I-56127 Pisa, Italy}

\date{\today}

\begin{abstract}

We investigate the particle-number dependence of some features of the
out-of-equilibrium dynamics of $d$-dimensional Fermi gases in the
dilute regime. We consider protocols entailing the variation of the
external potential which confines the particles within a limited
spatial region, in particular sudden changes of the trap size.  In
order to characterize the dynamic behavior of the Fermi gas, we
consider various global quantities such as the ground-state fidelity
for different trap sizes, the quantum work statistics associated with
the protocol considered, and the Loschmidt echo measuring the overlap
of the out-of-equilibrium quantum states with the initial ground
state.  Their asymptotic particle-number dependences show power laws
for noninteracting Fermi gases. We also discuss the effects of
short-ranged interactions to the power laws of the average work and
its square fluctuations, within the Hubbard model and its continuum
limit, arguing that they do not generally change the particle-number
power laws of the free Fermi gases, in any spatial dimensions.

\end{abstract}

\maketitle


\section{Introduction}
\label{intro}

The recent progress of experiments in atomic physics has provided a
great opportunity for a through investigation of the thermodynamics of
quantum systems, and the interplay between quantum and statistical
behaviors. Atomic systems are realized with a great control, thanks to
the impressive progress in the manipulation of cold
atoms~\cite{BDZ-08}.  The realization of physical systems which are
described by theoretical models, such as dilute Fermi and Bose gases,
Hubbard and Bose-Hubbard models, with different spatial dimensions
from one to three, provides through experimental checks of the
fundamental paradigma of statistical and quantum physics.  In
particular, they allow us to investigate the unitary quantum
evolution of closed many-body systems, exploiting their low
dissipation rate which maintains phase coherence for a long
time~\cite{BDZ-08,PSSV-11}.  Therefore the theoretical investigation
of the out-of-equilibrium unitary dynamics of many-body systems is of
great importance for a deep understanding of the fundamental issues of
quantum dynamics, their possible applications, and new developments.

In this paper we study some features of the out-of-equilibrium quantum
dynamics of Fermi gases, arising from variations of the external
potential which confines them within a limited spatial region.  We
consider generic $d$-dimensional traps arising from external power-law
potential, and in particular the cases of harmonic traps and hard-wall
traps.  Some aspects related to this issue have been discussed in the
literature, such as the time dependence of the particle density and
fixed-time correlation functions, spatial entanglement, etc..., in
particular for one-dimensional systems, see, e.g.,
Refs.~\cite{CGM-09,OS-02,PSOS-03,MG-05,RM-05,
  CM-06,GP-08,CV-10-BH,V-12-b,CSC-13,NV-13,CSC-13b,RBD-19}.
  
We focus on the particle-number dependence of the out-of-equilibrium
dynamics of $N$-particle Fermi gases in the dilute regime, when the
external potential is changed in such a way as to give rise to sudden
variations of the trap size, or shifts of the trap.  In order to
characterize the evolution of the quantum states, we consider various
global quantities, such as the ground-state fidelity associated
with changes of the trap size, the quantum work associated with a
sudden change of the trap size, the overlap between the quantum state
at a given time $t$ and the initial ground state as measured by the
so-called Loschmidt echo.  We show that large-$N$ power laws
characterize their dependence on the particle number.

We mostly consider lattice gas models of spinless noninteracting Fermi
particles in the dilute regime, realized in limit of large trap size
keeping the particle number fixed.  This corresponds to the trap-size
scaling limit, or continuum limit, whose scaling functions are related
to the correlation functions of a continuum many-body theory of free
Fermi particles in an external confining
potential~\cite{ACV-14,Nigro-17}. In the case of the quantum work and
its fluctuations, we also discuss the effects of particle
interactions, in the framework of the Hubbard model and its continuum
limit in the dilute regime.

The paper is organized as follows.  In Sec.~\ref{setting} we present
the general setting of the problem for free Fermi lattice gases in the
dilute regime, and their continuum limit.  In Sec.~\ref{fidelity} we
study the particle-number dependence of the ground-state fidelity
associated with variations of the trap size; the corresponding
equilibrium condition is realized in the limit of adiabatic changes of
the trap features.  Sec.~\ref{work} is devoted to the computation of
the first few moments of the quantum work distribution associated with
sudden changes of the trap size, starting for an equilibrium
(ground-state) condition.  In Sec.~\ref{overlap} we study the
particle-number dependence of the overlap between the quantum states
along the out-of-equilibrium evolution and the initial states, as
measured by the so-called Loschmidt echo.  In Sec.~\ref{ifs} we
discuss the effects of short-ranged particle interactions within the
Hubbard model and its continuum limit, arguing that the power laws of
the asymptotic particle-number dependence of the quantum work, and its
fluctuations, do not generally change with respect to the case of free
Fermi gases.  Finally, in Sec.~\ref{conclu} we summarize our main
results, and draw our conclusions.

\section{General setting of the problem}
\label{setting}

We consider $d$-dimensional lattice gases of $N$ noninteracting
spinless Fermi particles constrained within a limited spatial region
by an external force. The corresponding lattice many-body Hamiltonian
reads
\begin{eqnarray}
H(\ell) = - t \sum_{\langle {\bm x}{\bm y}\rangle} [c_{\bm x}^\dagger
c_{\bm y} + {\rm h.c.}] + \sum_{\bm x} V({\bm x},\ell) \,n_{\bm x}\,,
\label{freef}
\end{eqnarray}
where ${\bm x}$ are the sites of a $d$-dimensional cubic-like lattice,
$\langle {\bm x}{\bm y}\rangle$ indicates nearest-neighbor sites,
$c_{\bm x}$ is a spinless fermionic operator, $n_{\bm x}=c_{\bm
  x}^\dagger c_{\bm x}$ is the particle-density operator. In the rest
of the paper we set the lattice spacing $a=1$, the kinetic constant
$t=1$, and $\hslash = 1$; their dependence can be easily inferred by
dimensional analyses.  The confining potential $V({\bm x},\ell)$ is
coupled to the particle density operator; it is such that $V({\bm
  x},\ell)\to\infty$ for $|{\bm x}|\to\infty$, so that $\langle n_{\bm
  x}\rangle \to 0$ for ${\bm x}\to\infty$.  We assume it isotropic,
and characterized by a a generic power law, i.e.,
\begin{eqnarray}
  V({\bm x},\ell) = {1\over p} v^p |{\bm x}|^p\,, \quad \ell =
  v^{-1}\,,
\label{potential}
\end{eqnarray}
where $\ell$ should be considered as the trap
size~\cite{RM-05,ACV-14}.  The potential with power law $p=2$ gives
rise to harmonic traps, where $\omega=v$ is the corresponding
frequency.  In the limit $p\to\infty$ we recover hard-wall traps, so
that $V=0$ for $|{\bm x}|<\ell$ and $V=\infty$ for $|{\bm x}|>\ell$.
The particle number operator $\hat{N} = \sum_{\bm x}n_{\bm x}$ is
conserved, i.e., $[\hat{N},H(\ell)] = 0$. We consider the lattice
model (\ref{freef}) at a fixed number $N$ of particles,
$N\equiv\langle \hat{N}\rangle$.

We consider the dilute regime, when the particles are sufficiently
diluted, i.e., $N/\ell^d\ll 1$.  This is effectively defined as the
asymptotic behavior in the large trap-size limit, keeping the particle
number $N$ fixed.  This limit can be studied in the trap-size scaling
framework~\cite{CV-10-BH,ACV-14}, which relates the asymptotic
trap-size dependence of lattice gases in dilute regime with the
corresponding vacuum-to-metal quantum transition of the many-body
Hamiltonian (\ref{freef}) with a chemical potential term. We recall that the large
trap-size limit in the presence of a chemical potential $\mu$ [i.e.,
  adding a term $-\mu\sum_{\bm x} n_{\bm x}$ to the Hamiltonian
  (\ref{freef}), releasing the constraint on the number of
  particles] corresponds to taking the large-$\ell$ limit keeping the
ratio $N/\ell^d$ fixed. The {\em critical} behavior at the
vacuum-to-metal transitions (located at $\mu=\mu_c = - 2d$) is
characterized by the trap-size exponent~\cite{CV-10,ACV-14}
\begin{equation}
\theta = {p\over p+2} \,,
\label{thetaexp}
\end{equation}
depending on the power of the confining potential (\ref{potential}).
Its meaning is related to the fact the presence of an external
inhomogeneous potential induces a nontrivial length scale $\xi\sim
\ell^\theta$ in the correlation functions of the system.   Thus, the critical
length scale does not scale as the trap size, but as a nontrivial
power with exponent $\theta$. Only in the limit $p\to\infty$ we have
that $\xi\sim\ell$ as expected from standard finite-size scaling
arguments~\cite{CPV-14}.  For example the trap-size dependence of the
gap $\Delta(\ell)$ of the Fermi gas (i.e., the difference of the
lowest energy levels) behaves asymptotically as
\begin{equation}
\Delta({\ell}) \sim \xi^{-z} \sim \ell^{-z\theta},
\label{deltal}
\end{equation}
where $z=2$ is the dynamic exponent associated with the
vacuum-to-metal transition of Fermi gases. Moreover, correlation
functions of generic local operators ${\cal O}({\bm x})$ develop a
trap-size scaling behavior~\cite{CV-10-BH,ACV-14}, such as
\begin{eqnarray}
F({\bm x}_1,...,{\bm x}_n; \ell,N ) &\equiv& 
\langle {\cal O}({\bm x}_1) ... {\cal O}({\bm x}_n) \rangle 
\label{genf}\\
&\approx&
\ell^{-\varepsilon} 
{\cal F}({\bm X}_1,...,{\bm X}_n; N )
\nonumber 
\end{eqnarray}
where
\begin{equation}
{\bm X}_i={\bm x}_i/\ell^\theta,\qquad
\varepsilon = n\, \theta \,y_o,
\label{varepX}
\end{equation}
and $y_o$ is the renormalization-group dimension of the operator
${\cal O}({\bm x})$ at the fixed point associated with the
vacuum-to-metal transition~\cite{Sachdev-book,CPV-14}.  Of course,
corrections to this asymptotic behavior arise in lattice models, due
to the space discretization. They are generally suppressed by powers
of $\ell$, more precisely they are expect to vanish as
$\ell^{-2\theta}$ for lattice free-fermion gases.

In the {\em continuum} limit $a\to 0$, where $a$ is the lattice
spacing, or equivalently in the limit $\ell/a\to \infty$ keeping fixed
$a$, we recover a continuum model for a Fermi gas of $N$ particles in
a trap of size $\ell$, corresponding to the many-body problem with
one-particle Hamiltonian
\begin{eqnarray}
{\cal H}(\ell) = {{\bm p}^2\over 2m} + V({\bm
    x},\ell)\,.\label{oneph}
\end{eqnarray}
We set $m=1$, so that the trap size $\ell$ corresponds to that of the
lattice model (\ref{freef}), using the same unit ($\hslash=1$ and
$t=1$).   Such a continuum limit corresponds to the trap-size
scaling limit of the lattice model~\cite{CV-10,CV-10-BH}.  This
implies that the scaling functions ${\cal F}({\bm X}_1,...,{\bm X}_n; N )$
entering the trap-size scaling
relation (\ref{genf}) are exactly given by 
the continuum many-body
problem associated with the one-particle Hamiltonian (\ref{oneph}).
Some useful formulas for the ground state of Fermi gases with the
one-particle Hamiltonian (\ref{oneph}) are reported in
App.~\ref{groundstate}.

In this paper we mostly focus on the evolution of the Fermi gas
arising from variations of the trap size, starting from the ground
state associated with an initial trap size $\ell_0$.  We study the
relations between the initial and evolving states, as they are
quantified by a number of quantum-computing concepts, such as
ground-state fidelity, quantum work statistics, and Loschmidt echo.

In the protocol that we consider the initial condition of the Fermi
gas is the ground state associated with the initial Hamiltonian
parameters. Therefore, in the continuum limit, the $t=0$ state is
represented by the many-body wave function
\begin{equation}
\Psi({\bm x}_1,...,{\bm x}_N;t=0) = 
{1\over \sqrt{N!}} {\rm det} [\psi_i({\bm x}_j,\ell_0)]\,,
\label{fpsit0}
\end{equation}
where $\psi_k({\bm x},\ell_0)$ are the lowest $N$ eigenstates of the
one-particle Hamiltonian ${\cal H}(\ell_0)$, cf. Eq.~(\ref{oneph}).
Then the trapping potential generally changes as
\begin{equation}
V({\bm x},t) = {1\over p} \kappa(t) |{\bm x}|^p \,.
\label{vxt}
\end{equation}
The time dependence of the function $\kappa(t)$ has a time scale
$t_s$. In the limit $t_s\to 0$ we may consider it as a sudden change
of the confining potential, while for $t_s\to\infty$ we should recover
the adiabatic limit, when the quantum evolution passes through
equilibrium ground states associated with the varying trap sizes. The
time variation of the external potential gives generally rise to a
nontrivial quantum evolution of the Fermi gas, whose many-body wave
function in the continuum limit can be written a
\begin{equation}
\Psi({\bm x}_1,...,{\bm x}_N;t) = 
{1\over \sqrt{N!}} {\rm det} [\psi_i({\bm x}_j,t)]
\label{fpsit}
\end{equation}
where the one-particle wave functions $\psi_i({\bm x}_j,t)$ are
solutions of the one-particle Schr\"odinger problem
\begin{eqnarray}
&&i {d \psi_i({\bm x}_j,t)\over dt} = \left[
{{\bm p}^2\over 2} + V({\bm x},t)\right] \psi_i({\bm x}_j,t)\,,
\label{sctimed}\\
&&\psi_i({\bm x}_j,t=0) = \psi_i({\bm x},\ell_0)\,.
\label{incod}
\end{eqnarray}
In particular, we will consider the out-of-equilibrium dynamics
arising from sudden changes of the trap size.

\section{Ground-state fidelity related to variations of the trap size}
\label{fidelity}

Before discussing the out-of-equilibrium dynamics arising from sudden
variations of the trap size of the system, we 
investigate the adiabatic limit of our dynamic problem, 
which corresponds to slow variations
of the trap size $\ell(t)$, when the time scale of the time-dependent
external potential gets large, so that the system in always in the
ground state associated with the actual value $\ell(t)$.  Thus, the
global changes of the system properties are related to the variation
of the ground-state many-body wave function, and in particular to the
quantum overlap between the ground states for different trap sizes.
This is quantified by the equilibrium ground-state fidelity associated
with variations of the trap size.

The concept of ground-state fidelity has been introduced to quantify
the overlap between ground states associated with different parameters
of the model~\cite{Gu-10, BAB-17}.  The usefulness of the fidelity as
a tool to distinguish quantum states can be traced back to Anderson's
orthogonality catastrophe~\cite{Anderson-67}: the overlap of two
many-body ground states corresponding to Hamiltonians differing by a
small perturbation vanishes in the thermodynamic limit.

The ground-state fidelity monitors the changes of the ground-state
wave function $|0_{\ell,N}\rangle$ of the $N$-particle Fermi gas
trapped by the potential with length scale $\ell$, when varying the
control parameter $v=\ell^{-1}$.  We define it as~\cite{Gu-10}
\begin{equation}
F(\ell_0,\ell_1,N) \equiv | \langle 0_{\ell_1,N} |
0_{\ell_0,N} \rangle|\,.
\label{fiddef}
\end{equation}
Defining
\begin{equation}
\delta_\ell\equiv R_\ell -1\,,\qquad
 R_\ell\equiv \ell_1/\ell_0\,,
\label{deltaell}
\end{equation} 
and assuming $\delta_\ell$ sufficiently small, we can expand the
ground-state fidelity in powers of $\delta_\ell$:~\cite{Gu-10}
\begin{equation}
  F = 1 - \tfrac12 \delta_\ell^2 \, \chi_F(\ell_0,N) +
  O(\delta_\ell^3)\,,
  \label{expfide}
\end{equation}
where $\chi_F$ may be considered as the corresponding susceptibility.
The cancellation of the linear term in the expansion (\ref{expfide})
is essentially related to the fact that the fidelity is bounded, i.e.,
$0\le F\le 1$.  The fidelity susceptibility gives a quantitative idea
of the {\em speed} of the flow of ground states within the global
Hilbert space of the quantum states, when varying the trap size. 
The behavior of the
ground-state fidelity, and in particular its susceptibility, at
quantum transitions has been discussed in the literature, see, e.g.,
Refs.~\cite{RV-18,ZP-06,YLG-07,VZ-07}, finding a significant
enhancement with respect to the behavior of systems in normal
conditions.

We compute the ground-state fidelity in the trap-size scaling limit,
or equivalently in the continuum limit. As we shall see, the fidelity
susceptibility turns out to be independent of $\ell_0$ in this limit,
i.e., 
\begin{equation}
\chi_F(\ell_0,N)\equiv \chi_F(N)\,.\label{chifnind}
\end{equation}
We then determine the
large-$N$ asymptotic behaviors. It is important to note that such
large-$N$ asymptotic behaviors should be always intended within the
dilute regime of the lattice gas model, i.e., when the condition
$N/\ell^d\ll 1$ is satisfied.

To begin with, we consider $N$-particle Fermi gases constrained within
one-dimensional harmonic traps, whose ground-state wave function can
be written as~\cite{GWT-01}
\begin{eqnarray}
&&\Psi(x_1,...,x_N;\ell) 
= \ell^{-N/4} c_N A(X_1,...,X_N)
e^{-\sum_i X_i^2/2}\,,\nonumber \\
&&A(x_1,...,x_N) = \prod_{1\le i<j\le N}(X_i-X_j)\,,
 \label{phisolu} 
\nonumber
\end{eqnarray}
where $X_i=x_i/\sqrt{\ell}$, and $c_N$ is the appropriate
normalization constant so that $\int \prod_{i=1}^N dx_i
|\Psi|^2=1$.  The fidelity between one-dimensional ground states
associated with the trap sizes $\ell_0$ and $\ell_1$ can be
analytically computed, obtaining
\begin{eqnarray}
F(\ell_0,\ell_1,N) &=&
\int \prod_{i=1}^N dx_i \, 
\Psi(x_1,...,x_N;\ell_1)^* \Psi(x_1,...,x_N;\ell_0) \nonumber\\
&=& \left[ {4 \ell_0\ell_1 \over
    (\ell_0+\ell_1)^2} \right]^{N^2/4}\,.
\label{fl0l1}
\end{eqnarray}
By expanding it as in Eq.~(\ref{expfide}), we obtain the corresponding
susceptibility, which is given by
\begin{equation}
\chi_F(N)  = {1\over 8}\,N^2\,.
\label{chifhar}
\end{equation}

\begin{figure}[tbp]
\includegraphics*[scale=\graphicscale]{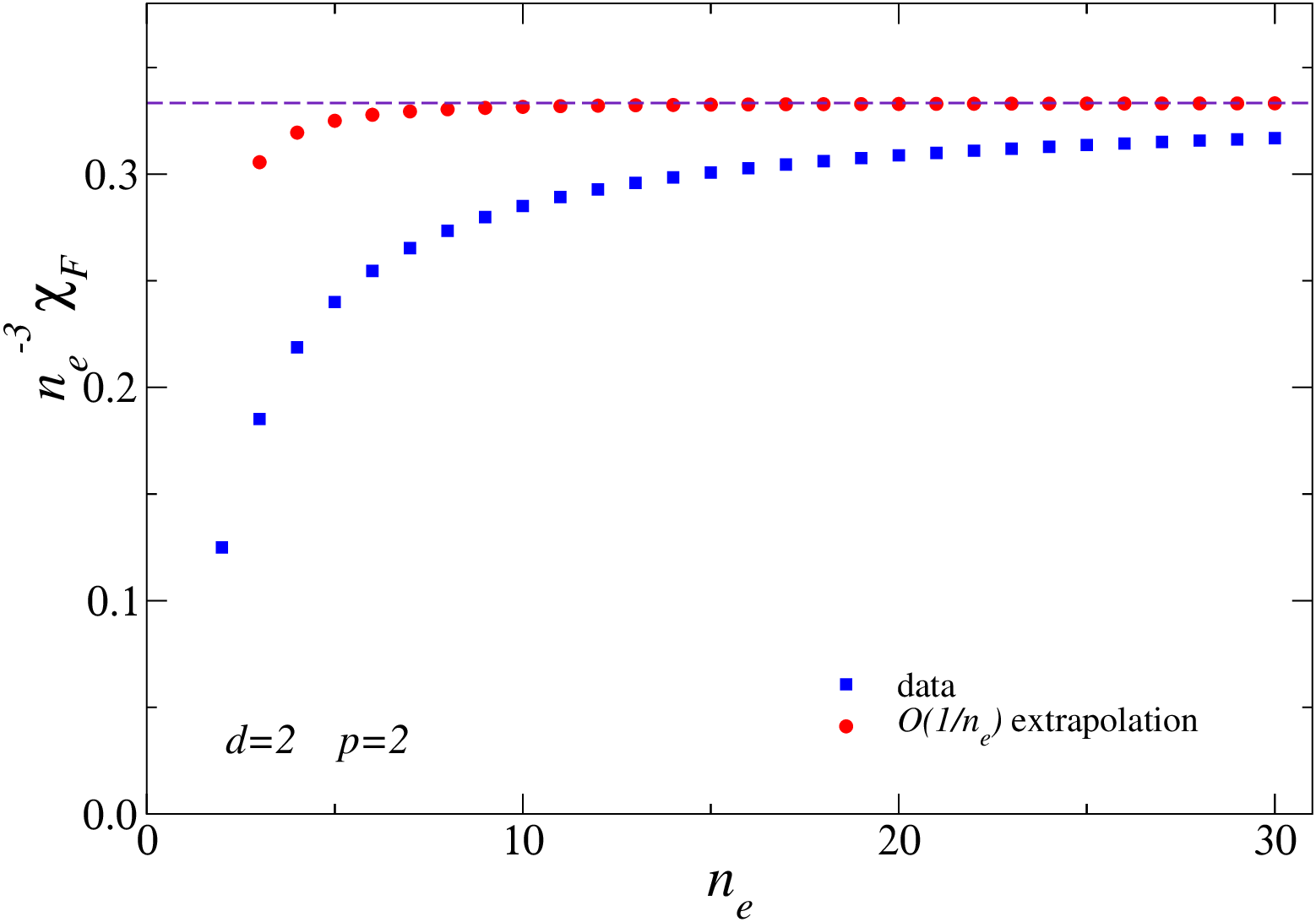}
\caption{ The fidelity susceptibility for two-dimensional harmonic
  traps with respect to a variation of the trap size $\ell$.  We show
  (practically exact) data of $n_e^{-3} \chi_F$ versus $n_e$, and also
  the corresponding linear extrapolation $a + b/n_e$ using the data
  for $n_e$ and $n_e-1$.  They clearly appear to approach the
  large-$n_e$ limit $n_e^{-3} \chi_F \approx 1/3$ shown by the dashed
  line.  Recalling that $n_e\approx \sqrt{2N}$ asymptotically, we
  obtain the large-$N$ behavior (\ref{chifharhd}).  }
\label{2dfide}
\end{figure}

The computation of the fidelity for Fermi gases in higher dimensions,
$d>1$, is more complicated.  The ground state of $N$-particle gases is
again given by the Slater determinant associated with the lowest $N$
one-particle states, such as Eq.~(\ref{fpsit0}).  They can be obtained
by filling all one-particle states (\ref{prodfunc}) with $\sum_i n_i
\le n_e$.  The number $N$ of particles/states is a function of $n_e$,
which asymptotically behaves as $N\approx n_e^2/2$ in two dimensions,
and $N\approx n_e^3/6$ in three dimensions.

The ground-state fidelity for different trap sizes is formally given
by integral of two $N$-particle Slater determinants.  To compute
matrix elements between states expressed in terms of Slater
determinants, such as $\Psi^{(\kappa)}({\bm x}_1,...,{\bm x}_N) = {\rm
  det} [\psi_i^{(\kappa)}({\bm x}_j)]/\sqrt{N!}$, we may use the
notable formula (see Ref.~\cite{Forrester-18} and references therein)
\begin{eqnarray}
&&\langle \Psi^{(1)}({\bm x}_1,...,{\bm x}_N) | \Psi^{(2)}({\bm
    x}_1,...,{\bm x}_N)\rangle =
\label{usfo}\\
&&=\int \prod_{i=1}^N d{\bm x}_i\; \Psi^{(1)}({\bm x}_1,...,{\bm
  x}_N)^* \Psi^{(2)}({\bm x}_1,...,{\bm x}_N) =\nonumber \\ 
&&= {\rm det}\left[\int d{\bm x} \,\psi^{(1)}_i({\bm x})^*
  \,\psi^{(2)}_j({\bm x}) \right]\,. \nonumber
\end{eqnarray}

We compute the fidelity associated with $N$ particles (in practice
this can be done exactly) by using the above formula with the
one-particle eigenfunctions associated with different trap sizes,
$\ell_0$ and $\ell_1$, so that $\delta_\ell\ll 1$.  Then, to evaluate
the fidelity susceptibility $\chi_F(N)$ for $N$ particles, we perform
the $\delta_\ell\to 0$ extrapolation of the quantity
$2(1-F)/\delta_\ell^2$ at fixed $N$. This can be achieved with high
accuracy.  Results for two-dimensional Fermi gases are shown in
Fig.~\ref{2dfide}.  The large-$N$ power law of $\chi_F$ is then
obtained by analyzing the behavior of the data with increasing $N$.
This analysis shows that the large-$N$ power law changes for harmonic
traps in higher dimensions. Indeed, the fidelity susceptibility shows
the asymptotic behavior
\begin{eqnarray}
\chi_F(N)  &=& b_d \, n_e^{d+1} \left[1 + O(n_e^{-1})\right] \nonumber\\
&=&  c_d \,N^{(d+1)/d} \left[ 1 + O(N^{-1/d})\right]\,,
\label{chifharhd}
\end{eqnarray}
for $d$-dimensional harmonic traps.  This is clearly supported by the
data shown in Fig.~\ref{2dfide} for two-dimensional gases up to
$n_e=30$ corresponding to $N=435$. We estimate $b_2 \approx 1/3$ with
high accuracy, see Fig.~\ref{2dfide}, thus $c_2\approx\sqrt{8/9}$.  An
analogous analysis of three-dimensional data confirms the large-$N$
behavior (\ref{chifharhd}) with $b_3 \approx 1/8$, thus $c_3 \approx
(81/32)^{1/3}$.

\begin{figure}[tbp]
\includegraphics*[scale=\graphicscale]{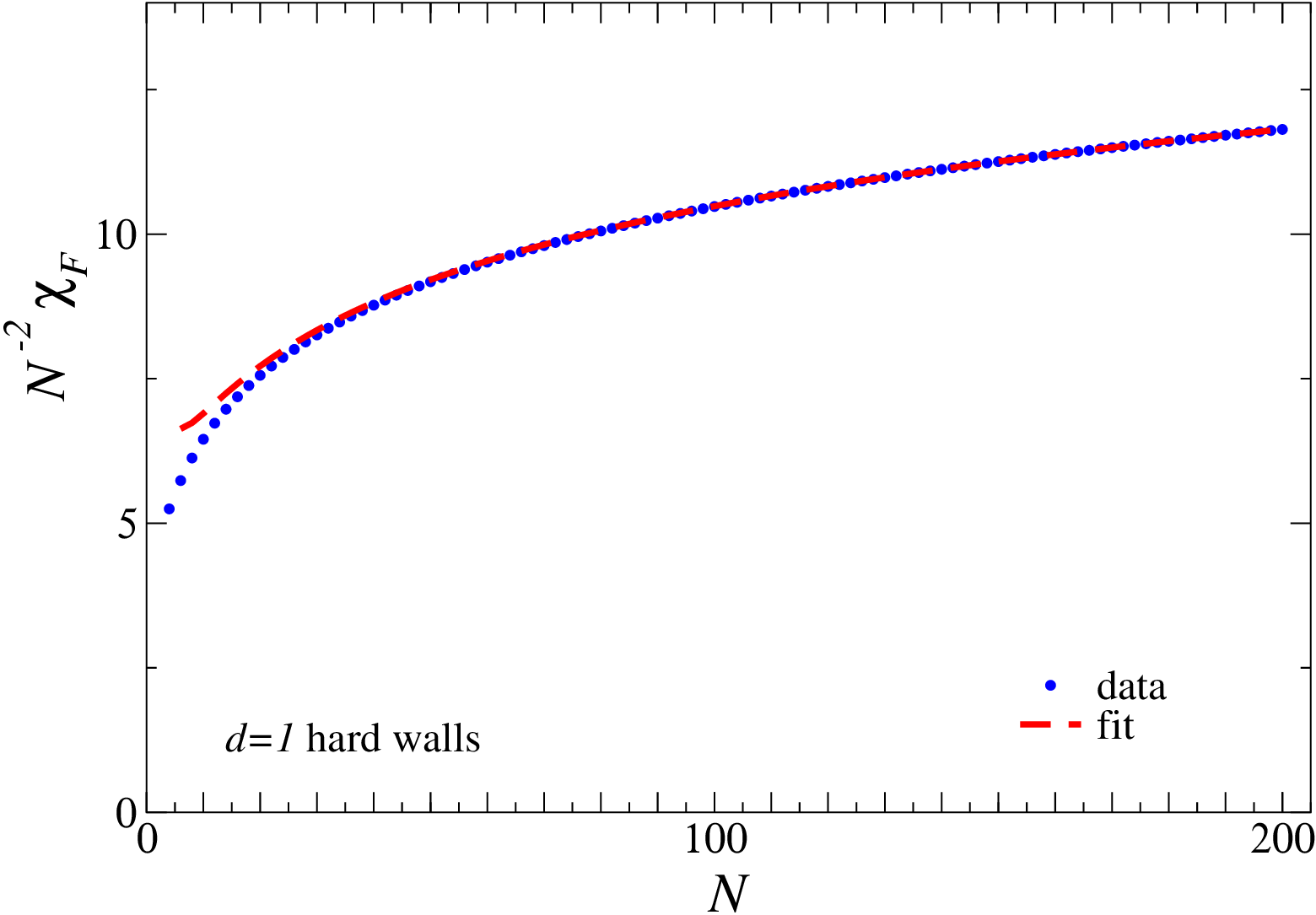}
\caption{ The fidelity susceptibility for one-dimensional hard-wall
  traps with respect to a variation of the trap size $\ell$.  We show
  data of $N^{-2} \chi_F$ up to $N=200$. Their large-$N$ behavior
  nicely fits the function $f(N)=2 \ln N + b + c/N$ as shown by the
  dashed line, supporting the asymptotic behavior (\ref{chiFhw}).  }
\label{1dfidepinf}
\end{figure}

We now consider the hard-wall limit
$p\to\infty$ of the confining potential.  In order to compute the
ground-state fidelity associated with two different trap sizes, we may
use the one-particle eigenfunctions (\ref{1deigf}) and the formula
(\ref{usfo}).  As shown by Fig.~\ref{1dfidepinf}, the results for
one-dimensional hard-wall traps show the asymptotic large-$N$
behavior
\begin{equation}
\chi_F(N) \approx a\, N^2 \ln N,
\label{chiFhw}
\end{equation}
with $a\approx 2$.  Therefore, it appears to increase faster than that
associated with the harmonic traps.

\section{Quantum work associated with changes of the trapping potential}
\label{work}

\subsection{Quantum work distribution}
\label{qdistr}

In this section we focus on the statistics of the work done on the
Fermi gas, when this is driven out of equilibrium by suddenly
switching the control parameter associated with the external
potential.  Several issues related to the definition and computation
of the work statistics in quantum systems have been already discussed
in a variety of physical implementations~\cite{CHT-11, GPGS-18},
including spin chains~\cite{Silva-08,DPK-08,Dorner-etal-12,
  Mascarenhas-etal-14, MS-14, ZT-15,SD-15,Bayat-etal-16,NRV-19},
fermionic and bosonic systems~\cite{DL-08,GS-12,SRH-14,
  SGLP-14,NRV-19}.
 
We consider the quantum dynamics of a ground-state Fermi gas initially
constrained within a trap of size $\ell_0$, that is subject to a
sudden variation of the trap size from $\ell_0$ to $\ell_1$.  In this
section we analyze the particle-number scaling of the quantum work
average and square fluctuations associated with this quench protocol.

The quantum work $W$ associated with out-of-equilibrium dynamic
protocols do not generally have a definite value.  More specifically,
this quantity can be defined as the difference of two projective
energy measurements~\cite{CHT-11}.  The first one at $t=0$ projects
onto the eigenstates of the initial Hamiltonian $H(\ell_0)$ with a
probability $p_{m,N}^{\ell_0}$ given by the density matrix of the
initial state, for example given by the equilibrium Gibbs
distribution. Then the system evolves, driven by the unitary operator
$U(t,0)=e^{-i H(\ell) t}$, and the second energy measurement projects
onto the eigenstates of the many-body Hamiltonian $H(\ell)$. The work
probability distribution can thus be written
as~\cite{CHT-11,TH-16,TLH-07}:
\begin{equation}
  P(W) 
  = \sum_{n,m} \delta \big[ W-(E_{n,N}^{\ell_1}-E_{m,N}^{\ell_0}) \big] \,
  \big| \langle n_{\ell_1,N} | m_{\ell_0,N} \rangle \big|^2 \,
  p_{m,N}^{\ell_0}\,,\quad
  \label{pwdefft}
\end{equation}
where $E_{n,N}^\ell$ and $|n_{\ell,N}\rangle$ are the eigenvalues and
corresponding eigenstates of the many-body Hamiltonian with trap size
$\ell$.  The zero-temperature limit corresponds to a quench protocol
starting from the ground state of $H(\ell_0)$ (we assume that the
ground-state is not degenerate).  The work probability~(\ref{pwdefft})
reduces to
\begin{equation}
  P(W) = \sum_{n} \delta \big[ W-(E_{n,N}^{\ell_1}-E_{0,N}^{\ell_0}) \big]
  \; \big| \langle n_{\ell_1,N} | 0_{\ell_0,N} \rangle \big|^2\,.
  \label{pwdef}
\end{equation}

Assuming that both $\ell_0$ and $\ell_1$ are large, thus in the
continuum or trap-size scaling limit, we conjecture that the work
probability develops the asymptotic behavior
\begin{eqnarray}
P(W,\ell_0,\ell_1,N) \approx \ell_0^{z\theta} {\cal
  P}(w,\delta_\ell,N)\,,\label{pwscabeh}
\end{eqnarray}
where we  have introduced the scaling variable
\begin{equation}
w = \ell_0^{z\theta} \,W\,,
\label{wdef}
\end{equation}
associated with the quantum work,
and $\delta_\ell=\ell_1/\ell_0-1$.
The power law of the prefactor of the work distribution and that of
the rescaling of the quantum work are related to the scaling behavior
of the gap, i.e. $\Delta(\ell_0)\sim \ell_0^{-z\theta}$, so that
\begin{equation}
\int dW\, P(W,\ell_0,\ell_1,N) = \int dw\,{\cal P}(w,\delta_\ell,N) = 1\,.
\label{normalization}
\end{equation}
The scaling behavior (\ref{pwscabeh}) implies that the moments
$\langle W^k \rangle$ of the work distribution develop the asymptotic
behavior
\begin{eqnarray}
&&\langle W^k \rangle=\int dW\,W\,P(W) \approx
\ell_0^{-z\theta k}\, {\cal W}_k(\delta_\ell,N)\,,
\quad\label{wkscal}
\end{eqnarray}
etc... These scaling relations will be supported by explicit
calculations.

We also mention that within the same scaling framework we may also
consider the more general case when the initial condition is
represented by a Gibbs distribution with temperature $T$, thus the
quantum work distribution is given by the more general expression
(\ref{pwdefft}), with $p_{m,N}^{\ell_0} \sim e^{-E_{m,N}^{\ell_0}/T}$.
For sufficiently small $T$, the temperature dependence can be taken
into account by adding a further scaling variable associated with $T$
to the arguments of the scaling functions. The corresponding scaling
variable is $T_r \sim T/\Delta(\ell_0)$ where $\Delta(\ell_0)\sim
\ell_0^{-z\theta}$ is the gap, cf. Eq.~(\ref{deltal}).
In the following we limit our calculations to the zero-temperature
limit.

\subsection{Average work}
\label{avwork}

Let us first determine the average work. We compute it in the
trap-size scaling or continuum limit.  Using Eqs.~(\ref{pwdef}) and
(\ref{wkscal}), we write it as
\begin{eqnarray}
\langle W\rangle  &=& 
\langle 0_{\ell_0,N} | \; H(\ell)-H(\ell_0)
\; | 0_{\ell_0,N}\rangle \label{avwo} \\ 
&=&\langle 0_{\ell_0,N} | \;\sum_{\bm x}\, 
[V({\bm x},\ell) - V({\bm x},\ell_0)] 
n_{\bm x}\; | 0_{\ell_0,N}\rangle \nonumber \\ 
&=& \int d{\bm x} [V({\bm x},\ell) - V({\bm x},\ell_0)] 
\rho({\bm x},\ell_0,N)\,,
\nonumber
\end{eqnarray}
where 
\begin{equation}
\rho({\bm x},\ell_0,N) = \langle 0_{\ell_0,N} | 
n({\bm x})  | 0_{\ell_0,N}\rangle\,.
\label{partdens}
\end{equation}
Therefore, the trap-size and particle-number dependences of the
average work can be inferred from those of the ground-state particle
density.  For $N$-particle Fermi gases, confined by a generic
power-law potential (\ref{potential}) with trap size $\ell_0$,
the trap-size scaling of the particle density can be obtained
from the corresponding continuum limit, i.e.,~\cite{V-12-a}
\begin{eqnarray} 
&&\rho({\bm x},\ell_0,N)\approx \ell_0^{-d\theta} S({\bm X},N)\,, 
\label{partdens2}\\
&&   {\bm X}\equiv {\bm x}/\ell_0^\theta\,,\qquad
S({\bm X},N) = \sum_{k=1}^N \psi_k({\bm X})^2\,,
\nonumber
\end{eqnarray}
where $\psi_k$ are the one-particle eigenfunctions of the one-particle
Hamiltonian (\ref{oneph}).  The large-$N$ behavior of $S_p$ turns out
to be~\cite{V-12-a,ACV-14}
\begin{equation}
S({\bm X},N) \approx N^\theta \, S_r({\bm X}/N^{(1-\theta)/d}).
\label{spxn}
\end{equation}
In particular, for a one-dimensional harmonic
trap~\cite{KB-02,GFF-05,CV-10-BH}
\begin{equation}
S_r(z) = {1\over \pi} \sqrt{2-z^2}\quad {\rm for} \;\;|z|\le
z_b=\sqrt{2}\,,
\label{p2pd}
\end{equation}
and $S_r(z)=0$ for $|z|\ge z_b$. 

Using Eq.~(\ref{partdens2}), we straightforwardly
obtain
\begin{eqnarray}
&&\langle W\rangle \approx \ell_0^{-2\theta} \,
{\cal W}_1(\delta_\ell,N)\,,\label{w1tss}\\
&&{\cal W}_1(\delta_\ell,N) = 
\,B(\delta_\ell)\,I_1(N)\,,\nonumber\\
&&B(x) = {1 - (1+x)^p\over p(1 + x)^p}= - x +O(x^2)\,,
\nonumber\\
&&I_1(N) =  \int d{\bm x} |{\bm x}|^p  S({\bm x},N)\,.
\nonumber
\end{eqnarray}
Note that this agrees with the trap-size scaling reported in Eq.~(\ref{wkscal}). 

Moreover, using Eq.~(\ref{spxn}), we obtain the asymptotic large-$N$
behavior
\begin{eqnarray}
&&{\cal W}_1(\delta_\ell,N) 
\approx 
B(\delta_\ell)\,
{\cal I}_1\, N^{1+2\theta/d}\,,
\label{largeNw1}\\
&&{\cal I}_1 = \int d{\bm x} |{\bm x}|^p  S_r({\bm x})\,.
\nonumber
\end{eqnarray}
Note that the above scaling equations imply first the trap-size
scaling limit, and then the large-$N$ limit, thus always remaining within
the dilute regime.

In particular, for one-dimensional harmonic traps, using
Eq.~(\ref{p2pd}),
\begin{eqnarray}
{\cal W}_1(\delta_\ell,N)  \approx 
{1\over 2}
\,B(\delta_\ell) \, N^{2}\,.
\label{largeNw1p2}
\end{eqnarray}
Note that, since  the ground-state energy is given by
\begin{equation}
E_0^{\ell} = \ell^{-1} \sum_{i=1}^N (i-1/2)= {N^2\over 2\ell}\,,
\label{e0ell}
\end{equation}
Eq.~(\ref{largeNw1p2}) implies
\begin{equation}
\langle W \rangle\ge E_0^{\ell_1} - E_0^{\ell_0} = - {1\over 2}
\ell_0^{-1} \,N^2\,{\delta_\ell\over 1 + \delta_\ell} \,,
\label{e0diff}
\end{equation}
as expected.

\subsection{Work fluctuations}
\label{workflu}

We now consider the second moment of the work distribution, and in
particular
\begin{equation}
\langle W^2\rangle_c = \langle W^2\rangle - \langle W\rangle^2\,.
\label{wconn}
\end{equation}
We obtain its scaling behavior, and in particular its large-$N$ power
law, by arguments similar to those for the average work.  
Using Eqs.~(\ref{pwdef}) and (\ref{wkscal}), we write
\begin{eqnarray}
&&\langle W^2\rangle_c = \langle 0_{\ell_0,N} | 
[H(\ell)-H(\ell_0)]^2 \;|0_{\ell_0,N}\rangle_c 
\label{aw2conn}\\
&&\;\; =\ell_0^{-2p} \,B(\delta_\ell)^2
\,\int d{\bm x}_1 d{\bm x}_2 |{\bm x}_1|^p |{\bm x}_2|^p 
G({\bm x}_1,{\bm x}_2) 
\nonumber
\end{eqnarray}
where 
\begin{eqnarray}
&&G({\bm x}_1,{\bm x}_2) =
\langle 0_{\ell_0,N} |\;
n({\bm x}_1) \,n({\bm x}_2) \;|
0_{\ell_0,N}\rangle \label{gxydef}\\
 &&\qquad -
\langle 0_{\ell_0,N} |\;
n({\bm x}_1) \;|0_{\ell_0,N}\rangle
\langle 0_{\ell_0,N} |\;
n({\bm x}_2) \;|0_{\ell_0,N}\rangle \,.
\nonumber
\end{eqnarray}
Therefore, the trap-size and particle-number dependences of the work
fluctuations can be inferred from those of the equilibrium
density-density connected correlation $G({\bm x}_1,{\bm x}_2)$.  

For free fermions, the following relations hold
\begin{equation}
G({\bm x}_1,{\bm x}_2) = 
- |C({\bm x}_1,{\bm x}_2)|^2 +
\delta({\bm x}_1-{\bm x}_2)  C({\bm x}_1,{\bm x}_2)\,,  
\label{gxycxy}
\end{equation}
where $C({\bm x}_1,{\bm x}_2)$ is the one-particle correlation 
function, which is~\cite{ACV-14}
\begin{eqnarray}
C({\bm x}_1,{\bm x}_2) &=& \langle 0_{\ell_0,N} |\;
c({\bm x}_1)^\dagger c({\bm x}_2)\;|0_{\ell_0,N}\rangle
\label{cxy}\\
&=& \ell_0^{-d\theta}\,E({\bm X}_1,{\bm X}_2)\,,
\nonumber 
\end{eqnarray}
where
\begin{eqnarray}
E({\bm X}_1,{\bm X}_2) = 
\sum_{k=1}^N \psi_k({\bm X}_1)^*
\psi_k({\bm X}_2)\,,
\quad {\bm X}_i = {\bm x}_i/\ell_0^{\theta}\,.\quad
\label{epexr}
\end{eqnarray}
Of course, $C({\bm x},{\bm x}) = \rho({\bm x})$.  
Therefore, we have
\begin{eqnarray}
&&G({\bm x}_1,{\bm x}_2) = \ell_0^{-2d\theta} Z({\bm X}_1,{\bm X}_2)\,,
\label{gxy}\\
&&Z({\bm X}_1,{\bm X}_2) = 
-E({\bm X}_1,{\bm X}_2)^2 + \delta({\bm X}_1-{\bm X}_2)
E({\bm X}_1,{\bm X}_2)\,.
\nonumber 
\end{eqnarray}
The trap-size scaling function $Z$ 
develops the large-$N$ behavior~\cite{V-12-a,ACV-14}
\begin{equation}
Z({\bm X}_1,{\bm X}_2,N) 
\approx N^\theta \, Z_r(N^{\theta/d}{\bm X}_1,
N^{\theta/d}{\bm X}_2)
\label{epxn}
\end{equation}
for ${\bm X}_1 \neq {\bm X}_2$ (this scaling behavior does
not hold for $|{\bm X}_1-{\bm X}_2|\to 0$).

\begin{figure}[tbp]
\includegraphics*[scale=\graphicscale]{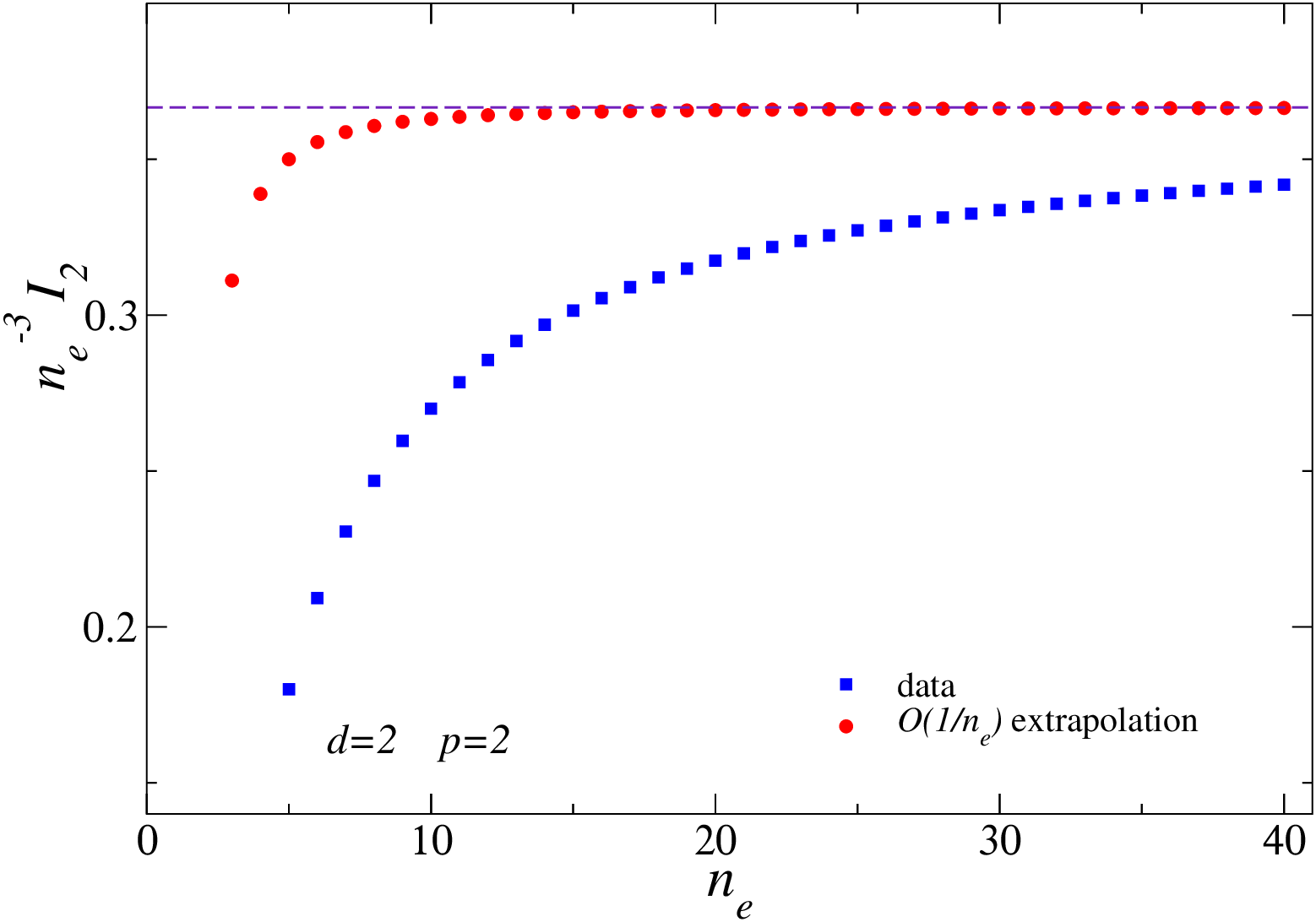}
\caption{$N$-dependence of the function $I_2(N)$,
  cf. Eq.~(\ref{jpnexp}), of the square work fluctuations $\langle
  W^2\rangle_c$, cf. Eq.~(\ref{tssw2}), associated with a sudden
  quench of the trap size of two-dimensional gases, from $\ell_0$ to
  $\ell_1$. We show data of $n_e^{-3} I_2$ versus $n_e$, and also the
  corresponding linear extrapolation $a + b/n_e$ using the data for
  $n_e$ and $n_e-1$.  They approach the large-$n_e$ limit $n_e^{-3}I_2
  \approx 1/3$ shown by the dashed line.  Recalling that
  asymptotically $n_e\approx \sqrt{2N}$, we obtain the large-$N$
  behavior (\ref{Jp2dp2}).  }
\label{2dw2}
\end{figure}

Using the above results for the connected density-density correlation
function, we arrive at
\begin{eqnarray}
&&\langle W^2\rangle_c \approx \ell_0^{-4\theta}\, {\cal W}_2(\delta_\ell,N)\,,
\label{tssw2}\\
&&
{\cal W}_2(\delta_\ell,N) = B(\delta_\ell)^2 \, I_2(N)\,,
\nonumber\\
&&I_2(N) = \int d{\bm X}_1 d{\bm X}_2 |{\bm X}_1|^p |{\bm X}_2|^p 
Z({\bm X}_1,{\bm X}_2) \,.
\nonumber
\end{eqnarray}
Again, Eq.~(\ref{tssw2}) agrees with the trap-size scaling put forward
in Eq.~(\ref{wkscal}).

The large-$N$ dependence can hardly be inferred from the large-$N$
scaling of the two-point function $G({\bm x}_1,{\bm x}_2)$, such as
Eq.~(\ref{epxn}), because the integral $I_2$ in Eq.~(\ref{tssw2}) includes the
contribution for $|{\bm x}_2-{\bm x}_1|\to 0$, where Eq.~(\ref{epxn})
does not apply.  In order to determine it, we compute
\begin{eqnarray}
&&I_2(N) = {\rm Tr} M_{2p} - {\rm Tr} M_{1p}^\dagger M_{1p}
  \,,\label{jpnA}\\ &&M_{kp,ij} = \int d{\bm x} |{\bm x}|^{kp}
  \psi_i({\bm x})^* \psi_j({\bm x}) \,.
\label{akp}
\end{eqnarray}
The analysis of $I_2(N)$ with increasing $N$ shows that the leading
contributions of the two terms in Eq.~(\ref{jpnA}) asymptotically
cancel.  For one-dimensional harmonic traps we obtain the exact result
\begin{eqnarray}
I_2(N) = {1\over 2} N^2 \quad {\rm for}\;d=1,\;\;p=2\,.\quad
\label{Jp1dp2}
\end{eqnarray}
For two-dimensional harmonic traps, the large-$N$ extrapolation of
fixed-$N$ results shows the asymptotic behavior
\begin{eqnarray}
I_2(N) = e N^{3/2} \left[ 1 + O(N^{-1/2})\right]
\quad {\rm for}\;d=2,\;\;p=2\,,\quad
\label{Jp2dp2}
\end{eqnarray}
with $e \approx \sqrt{8/9} $, see Fig.~\ref{2dw2}.  These results may
hint at the general large-$N$  behavior $I_2(N)
\approx e \,N^{1 + 2\theta/d}$ when extending the results to confining potential
with generic powers $p$.

We finally note that the moments of the work distribution
(\ref{pwdef}) cannot always written as expectation values of powers of
the difference of the Hamiltonians, such as the cases of the first and
second moment, cf.  Eqs.~(\ref{avwo}) and (\ref{aw2conn}).  For higher
moments more complicated expressions must be evaluated.  In this paper
we only report results for the first two moments.

\subsection{Quantum work associated with the Fermi pendulum}
\label{movcen}

Let us now consider a noninteracting one-dimensional Fermi gas of $N$
particles trapped by a harmonic potential.  The initial state at $t=0$
is its ground state within a trap of size $\ell_0\equiv\omega_0^{-1}$
centered at a distance $x_c$: $\Psi(x,0) = {1\over \sqrt{N!}}{\rm det}
[\psi_i(x_j-x_c)]$ where $\psi_i$ are the one-particle eigenstates in
a harmonic potential.  Then the gas is released within a larger trap
of size $\ell\equiv\omega^{-1}>\ell_0$.  The time-dependent many-body
function describing the motion is given by
\begin{eqnarray}
&&\Psi(x_1,...,x_N,t) = {1\over \sqrt{N!}}{\rm det} [\psi_i(x_j,t)],
\label{psixmt} \\
&&\psi_i(x,t) = \int_{-\infty}^{\infty} dy P(x,t;y,0)
  \phi_i(y-x_c,\ell_0), \nonumber \\
&&P(x,t;y,0) = \left[{m\omega \over 2\pi i \sin(\omega
    t)}\right]^{1/2}  \times \nonumber\\
&&\quad    \times
    \exp\left\{ {im\omega\over 2\sin(\omega
      t)}[(x^2+y^2)\cos(\omega t) - 2xy]\right\}\,, \nonumber
\end{eqnarray}
where $ \phi_i(y-x_c,\ell_0)$ are the eigenfunctions (\ref{1deigf})
for a harmonic trap centered at $x_0$ with trap size $\ell_0$.  The
particle density oscillates as a pendulum, as one can easily
infer by computing the time dependence of the particle density,
\begin{equation}
\rho(x,t) = \sum_{i=1}^N |\psi_i(x,t)|^2 = \rho(x,t+2\pi/\omega)\,,
\label{rhoxt}
\end{equation}
and $\rho(x,t) = \rho(-x,t+\pi/\omega)$.

We now compute the average quantum work associated with the sudden
shift, the center moving from $x=x_c$ to $x=0$, 
and enlargement of the harmonic trap. For this purpose we must
evaluate
\begin{eqnarray}
\langle W \rangle &=&
\langle 0_{x_c,\ell_0,N} \,|\,
{1\over 2\ell_1^2} \int dx\,  x^2 n(x) 
| \,0_{x_c,\ell_0,N} \rangle
\label{avwshift}\\
&-&
\langle 0_{x_c,\ell_0,N} \,|\,
{1\over 2\ell_0^2} \int dx\,  (x-x_c)^2 n(x) \,
| \,0_{x_c,\ell_0,N} \rangle\,.
\nonumber
\end{eqnarray}
where $| \,0_{x_c,\ell_0,N} \rangle$ indicates the ground state of the
$N$-particle Fermi gas in a harmonic trap of size $\ell_0$ centered at
$x_c$.  After some manipulations, we may write it as
\begin{eqnarray}
\langle W \rangle &=& \ell_0^{-2} \,
\langle 0_{x_c,\ell_0,N} \,|\, 
B(\delta_\ell) \int dx\, (x-x_c)^2 n(x) 
\,| \,0_{x_c,\ell_0,N} \rangle 
\nonumber \\
&+&\ell_0^{-2} \,
\langle 0_{x_c,\ell_0,N} \,|\, 
 {x_c\over R_\ell^{2}} \int dx\, (x-x_c)n(x)    
\,| \,0_{x_c,\ell_0,N} \rangle \nonumber \\
&+&\ell_0^{-2} \,\langle 0_{x_c,\ell_0,N} \,|\, 
{x_c^2\over 2 R_\ell^2} \int dx\, n(x) 
\,| \,0_{x_c,\ell_0,N} \rangle\,,
\label{calcwsh}
\end{eqnarray}
where $R_\ell=\ell_1/\ell_0$. 
Thus
\begin{eqnarray}
\langle W \rangle &=& \ell_0^{-2} B(\delta_\ell) \int dx\, (x-x_c)^2
\,\rho(x-x_c,\ell_0) \label{calcwsh2}\\ &+& \ell_0^{-2} {x_c\over
  R_\ell^{2}} \int dx\,(x-x_c)\,\rho(x-x_c,\ell_0)
\nonumber\\ &+&\ell_0^{-2}{x_c^2\over 2 R_\ell^2} \int dx\,
\rho(x-x_c,\ell_0)\,, \nonumber
\end{eqnarray}
where $\rho(x,\ell_0)$ is the particle density for a trap size
$\ell_0$. 
We note that the first term corresponds to the average work
for the variation of the trap size from $\ell_0$ to $\ell_1$,
cf. Eq.~(\ref{avwo}), and the second one vanishes because of the
reflection symmetry of the particle density.  We obtain
\begin{eqnarray}
&&\langle W \rangle =\ell_0^{-1}\,{\cal W}_1(\delta_\ell,X_c,N)\,,
\label{wshift}\\
&&{\cal W}_1(\delta_\ell,X_c,N) = 
{B(\delta_\ell)\over 2} N^2 +
  {X_c^2\over 2 R_\ell^2} N\,,
\nonumber
\end{eqnarray}
where $X_c=x_c/\ell^\theta$.  The first term is essentially related to
the change of the trap size, while the second one to the shift of the
trap. Note that their $N$-dependence power law differs; the dominant
one is that related to the change of the trap size.

We may also compute the average fluctuations $\langle W^2 \rangle_c$.
We only report the results for the case
the trap size is unchanged, thus $\ell_1=\ell_0$, and we only 
shift the trap center from $x_c$ to the origin.  We obtain
\begin{eqnarray}
&&\langle W^2 \rangle_c = \ell_0^{-2}\,{\cal W}_2(\delta_\ell,X_c,N)\,,
\label{w2sh}\\ 
&&{\cal W}_2 = X_c^2\, \int dX_1\, dX_2 \,X_1 \,X_2 \,Z(X_1,X_2)
=X_c^2 {N\over 2}\,.  \nonumber
\end{eqnarray}

\section{Quantum overlap between initial and evolved states}
\label{overlap}

\subsection{The Loschmidt echo}
\label{loscecho}

In order to characterize the quantum dynamics arising from variations
of the trapping potential, we study how the out-of-equilibrium states
arising from the change of the trapping potential depart from the
initial one, which is the ground state associated with the trap size
$\ell_0$. This issue can be quantitatively analyzed by considering the
{\em overlap} between the initial state and the evolving $N$-particle
states during the out-of-equilibrium quantum evolution.  
This provides nontrivial
information on the nature of the quantum dynamics associated with
the quenches considered in this paper, extending earlier
studies focussing on the correlation functions and spatial
entanglement at fixed time~\cite{V-12-b,NV-13}.

The evolution of the overlap with the initial ground state can be quantified
by the so-called Loschmidt echo,
\begin{eqnarray}
L_E = | \langle \Psi({\bm x}_1,...,{\bm
  x}_N;t) | \Psi({\bm x}_1,...,{\bm x}_N;t=0) \rangle|\,,\quad
\label{lechoq}
\end{eqnarray}
and the related echo function
\begin{eqnarray}
Q(t,N) = - \ln L_E(t,N)\;, \label{qdef}
\end{eqnarray}
where the initial $t=0$ state $| \Psi({\bm x}_1,...,{\bm x}_N;t=0)
\rangle$ is the ground state for system constrained within a trap of
size $\ell_0$. Therefore, the echo function $Q$ becomes larger and
larger when the overlap measured by the Loschmidt echo 
gets more and more suppressed.

We consider again a sudden quench of the potential, corresponding to the
variation of the trap size from $\ell_0$ to $\ell_1$, including
$\ell_1\to\infty$ corresponding to a free expansion of the gas. We
generally expect the following scaling behavior
\begin{equation}
Q(t,\ell_0,\ell_1,N) \approx {\cal Q}(\tau,\delta_\ell,N)\,,
\label{qtscaN}
\end{equation}
where 
\begin{equation}
\tau = \ell_0^{-z\theta} t\,
\label{thetadef}
\end{equation}
is a scaling variable associated with the time $t$,
so that $\tau\sim t\,\Delta({\ell_0})$, since
$\Delta({\ell_0})\sim\ell_0^{-z\theta}$ is the gap for the trap of
size $\ell_0$. The dynamic trap-size scaling behavior (\ref{qtscaN})
is analogous to that put forward, and numerically checked,
for the dynamic finite-size scaling of the
Loschmidt echo in out-of-equilibrium conditions 
arising from quenches at quantum
transitions~\cite{PRV-18}.

In the following we focus on one-dimensional systems trapped by
harmonic and hard-wall potentials. Extensions to higher dimensions can
be straightforwardly considered, but require more cumbersome
calculations. We present calculations in the continuum limit, which
are valid in the trap-size scaling limit of the lattice gas model.

\subsection{Harmonic traps}
\label{dynhartraps}

\subsubsection{Quantum dynamics when changing one-dimensional harmonic traps}
\label{tddm}

We consider Fermi gases in general time-dependent confining harmonic
potential, Eq.~(\ref{vxt}) with $p=2$, starting from an equilibrium
ground state configuration with initial trap size $\ell_0$, as
outlined in Sec.~\ref{setting}.

As shown in Ref.~\cite{KSS-96}, see also \cite{MG-05}, the
time-dependent many-body wave function 
$\Psi(x_1,...,x_N;t)$ of the system can be derived from the
solutions $\psi_j(x,t)$ of the one-particle Schr\"odinger equation
\begin{eqnarray}
i\partial_t \psi_j(x,t) = \left[ -{1\over 2}\partial_x^2 + {1\over 2}
  \kappa(t) x^2\right] \psi_j(x,t),\label{scoeq}
\end{eqnarray}
with the initial condition $\psi_j(x,0)=\phi_j(x)$ where $\phi_j(x)$
are the eigensolutions of the Hamiltonian at $t=0$, characterized by a
trap size $l_0$, with eigenvalue $E_j = \ell_0^{-1} (j-1/2)$.  The
solution can be obtained introducing a time-dependent function $s(t)$,
writing~\cite{PP-70,KSS-96}
\begin{eqnarray}
\psi_j(x,t) = && s^{-1/2} \phi_j(x/s) \times \label{psijsol}\\
&&\times {\rm exp}\left( 
i {\dot{s}x^2\over 2 s} - i E_j\int_0^t s^{-2} dt' \right),
\nonumber
\end{eqnarray}
where $\phi_j(x)$ is the $j^{\rm th}$ eigenfunction of the
Schr\"odinger equation of the one-particle Hamiltonian at $t=0$, thus
with trap size $l_0$. The function $s(t)$ satisfies the nonlinear
differential equation
\begin{equation}
\ddot{s} + \kappa(t) s = \kappa_0 s^{-3}
\label{fdef}
\end{equation}
with initial conditions $s(0)=1$ and $\dot{s}(0)=0$.  Then, using
Eq.~(\ref{psijsol}), one can write the time-dependent 
many-body wave function as~\cite{MG-05}
\begin{eqnarray}
&&\Psi(x_1,...,x_N;t) =  {1\over \sqrt{N!}} {\rm det} [\psi_j(x_i,t)]
\nonumber \\
&&\quad =s^{-N/2}\Psi(x_1/s,...,x_N/s;0)\times
\nonumber \\
&&\quad\times \;{\rm exp}\left( {i\dot{s}\over 2s}\sum_j x^2_j -
i \sum_j E_j \int_0^t s^{-2} dt' \right),
\label{phisoldyn}
\end{eqnarray}
where $\Psi(x_1,...,x_N;0)$ is the wave function of the ground state
for the Hamiltonian at $t=0$.

In the case of an instantaneous change to a confining potential with
trap size $\ell_1$, so that $\kappa(t)=\ell_1^{-2}$ for $t>0$, the
solution of Eq.~(\ref{fdef}) reads
\begin{equation}
s(t) = \sqrt{ 1 + (R_\ell^2-1) \left[{\rm sin}(t/(R_\ell \ell_0)\right]^2}\,,
\label{stsol}
\end{equation}
where $R_\ell=\ell_1/\ell_0$.  Notice that, assuming $R_\ell>1$, 
\begin{equation}
1\le s(t)\le R_\ell\,,
\label{stinterv}
\end{equation}
and $\dot{s}=0$ when $s(t)=1$ and $s(t)=R_\ell$.  Interestingly, the
many-body quantum states at times corresponding to $s(t)=1$ and
$s(t)=R_\ell$ turn out to coincide with the ground states associated
with the trap sizes $\ell=\ell_0$ and $\ell=\ell_0
R_\ell^2=\ell_1^2/\ell_0$ respectively.  In
the case $\ell_1\to\infty$, corresponding to an instantaneous drop of
the trap, so that $\kappa(t)=0$ for $t>0$, the solution of
Eq.~(\ref{fdef}) is a monotonically increasing function, given by
\begin{equation}
s(t) = \sqrt{1+(t/\ell_0)^2}.
\label{ftqinf}
\end{equation}
Further analytic results for a linear time dependence of $\kappa(t)$
in Eq.~(\ref{vxt}) can be found in Ref.~\cite{CV-10-BH}.

\subsubsection{The Loschmidt echo for an instantaneous change
of the trap size}
\label{lechtrap}

Using the above results, we may write the Loschmidt echo as
\begin{eqnarray}
&&L_E(t,N) = | \;s^{-N/2} \int \prod_{i=1}^N dx_i\;
  \Psi(x_i/s;0)^*\Psi(x_i;0) \times\nonumber\\ &&\qquad
  \qquad\qquad\times \;{\rm exp}\left( {-i\dot{s}\over 2s}\sum_j
  x^2_j\right) |
\label{eltn}\\
&&\;= 
s^{-N/2} | {\rm det} \int dx \,\psi_i(x/s;0)^*\psi_j(x;0) 
{\rm exp}\left( {-i\dot{s}\over 2s}x^2\right) |\,,
\nonumber
\end{eqnarray}
where we used Eq.~(\ref{usfo}).
Then, noting that the function $s(t)$, cf. Eq.~(\ref{stsol}), can be
rewritten as
\begin{equation}
s(t)\equiv S(\tau,\delta_\ell) \,,
\label{stsca}
\end{equation}
where $\tau = t/\ell_0$,
$\delta_\ell = \ell_1/\ell_0-1$, and
\begin{eqnarray}
&&S(\tau,\delta_\ell) = \left[1 + (2\delta_\ell + \delta_\ell^2)
    \left(\sin{\tau\over 1+\delta_\ell}\right)^2\right]^{1/2}\;,
  \qquad \label{stheta}\\ 
  &&S(\tau,\delta_\ell=\infty) =
  \sqrt{1+\tau^2}\,,
\nonumber
\end{eqnarray}
we obtain
\begin{eqnarray}
&&L_E(t,\ell_0,\ell_1,N) = | {\rm det} A_{ij}(\tau,\delta_\ell,N) |\,,
\label{eltn3}\\
&&A_{ij} = \int dZ e^{-i S' Z^2/2} \phi_i(Z/\sqrt{S}) \phi_j(Z\sqrt{S})\,,
\nonumber
\end{eqnarray}
where $S'= dS/d\tau$, and the eigenfunctions $\phi_n(X)$ 
are those reported in 
Eq.~(\ref{1deigf}) with $\xi=1$.

By further developing the above equations, we arrive at the final
expression
\begin{eqnarray}
&&L_E(t,\ell_0,\ell_1,N)  = F(S,S')^{N^2/2}\,,
\label{finalq}\\
&&F(S,S') = {2 S\over \sqrt{(1+S^2)^2 + S^2 S'^2}}\,.
\nonumber
\end{eqnarray}
This can be derived by straightforward manipulations of
the expression (\ref{eltn}), or by exploiting the properties of the
Hermite polynomials entering the determinant (\ref{eltn3}).
We have also checked it numerically.  

\begin{figure}[tbp]
\includegraphics*[scale=\graphicscale]{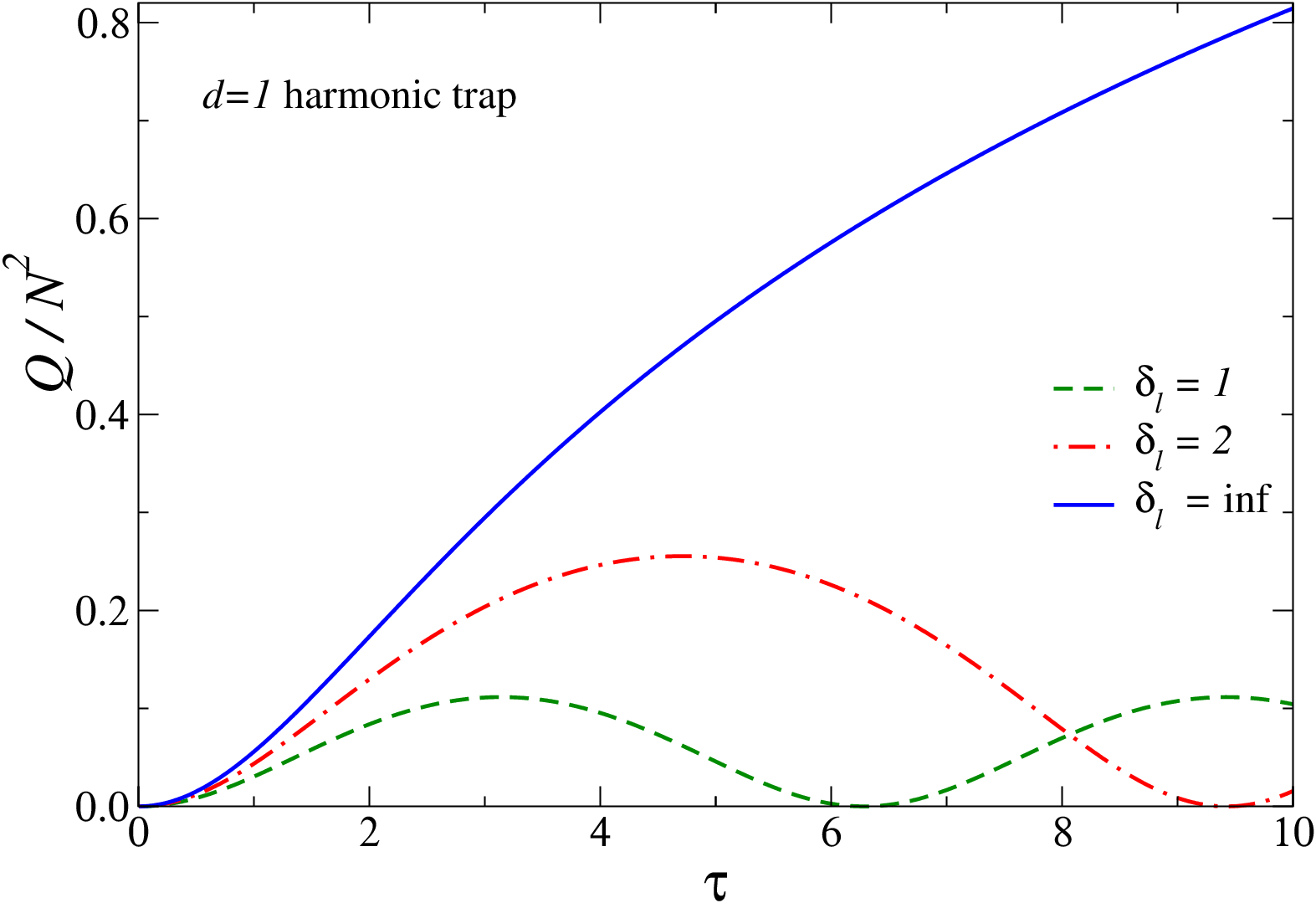}
\caption{ The echo function $Q$, cf. Eqs.~(\ref{lechoq}) and
  (\ref{qdef}), associated with changes of one-dimensional harmonic
  traps, whose trap size suddenly varies from $\ell_0$ to $\ell_1$,
  for $\delta_\ell=1,\,2,\,\infty$, as given by Eq.~(\ref{lochecho}).
}
\label{hatrapLe}
\end{figure}

Finally, for the echo function $Q = -\ln L_E$ we obtain
\begin{equation}
Q(t,\ell_0,\ell_1,N) = 
{N^2\over 4} \ln\left[ {(1+S^2)^2 +
    S^2 S'^2\over 4 S^2}\right]\,,
\label{lochecho}
\end{equation}
where $S(\tau,\delta_\ell)$ is reported in Eq.~(\ref{stheta}).  The
above expression is in agreement with the general scaling behavior put
forward in Eq.~(\ref{qtscaN}).  In Fig.~\ref{hatrapLe} we show the
echo function for some values of $\delta_\ell$, including that for the
free expansion $\delta_\ell\to\infty$.  Note that when $Q(t)=0$, the
quantum state coincides with the initial one, apart from a trivial
phase; this occurs periodically, when $\tau = k \pi R_\ell$ and
$k=0,1,2,...$.  In the case of a free expansion, $R_\ell=\infty$, we
have
\begin{equation}
Q(t,\ell_0,\infty,N) \approx {N^2\over 2}\ln\tau
\label{lochechoinf}
\end{equation}
in the large-time limit.

\subsection{Free expansion from a hard-wall trap}
\label{freexphw}

We now consider an $N$-particle Fermi gas constrained within 
hard walls, in the corresponding ground state,  and study the
out-of-equilibrium dynamics arising from the sudden drop
of the hard walls, 
allowing the Fermi gas to expand freely.

\begin{figure}[tbp]
\includegraphics*[scale=\graphicscale]{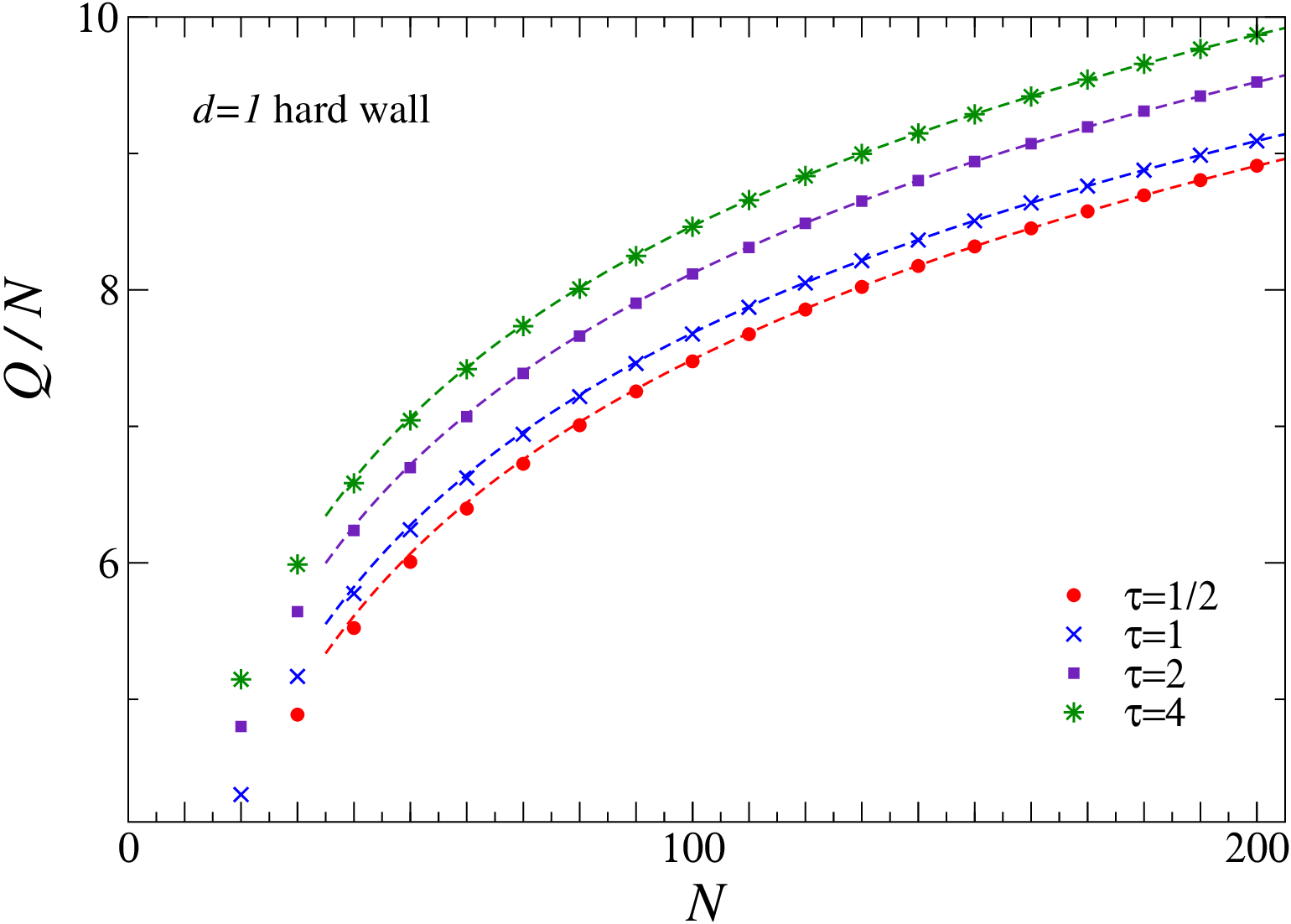}
\caption{ The echo function $Q$, cf. Eqs.~(\ref{lechoq}) and
  (\ref{qdef}), for a free expansion of the Fermi gas after the sudden
  drop of the hard walls trapping the gas, for some values of
  $\tau=\ell_0^{-2} t$.  The data are consistent with the asymptotic
  behavior $Q \approx a N (\ln N + b)$ (represented by the dashed
  lines).  }
\label{hdquench}
\end{figure}

The free expansion of the gas after the instantaneous drop of the
walls is described by the time-dependent wave function
\begin{equation}
\Psi(x_1,...,x_N;t) = {1\over \sqrt{N!}} {\rm det}  [\psi_i(x_j,t)]
\label{fpsi}
\end{equation}
where $\psi_i(x,t)$ are the one-particle wave functions with initial
condition $\psi_i(x,0)=\phi_i(x)$, cf. Eq.~(\ref{1deigfinf}), which
can be written in terms of the free-particle propagator
$P(x_2,t_2;x_1,t_1)$, as
\begin{eqnarray}
&&\psi_i(x,t) = \int_{-\ell_0}^{\ell_0} dy \,P(x,t;y,0)
  \,\phi_i(y)\,,\label{fexev}\\ &&P(x_2,t_2;x_1,t_1)= {1\over \sqrt{i
      2\pi (t_2-t_1)} } \exp\left[{i(x_2-x_1)^2\over
      2(t_2-t_1)}\right]\,.  \nonumber
\end{eqnarray}

Then the Loschmidt echo can be written as
\begin{eqnarray}
L_E(\ell_0,t,N) = | \langle \Psi_0 | \Psi(t) \rangle | = 
| {\rm det} F_{kq}(t) |\,,\quad
\label{lechfkq}
\end{eqnarray}
where 
\begin{eqnarray}
F_{kq}(t) = \int_{-\ell_0}^{\ell_0} dx\, \phi_k(x)^*
\,\psi_q(x,t)\,. \label{fkq}
\end{eqnarray}
One can easily check that $L_E$, thus $Q=-\ln L_E$, can be written as
a function of the scaling variable $\tau=\ell_0^{-2} t$ and the
particle number $N$, in agreement with the scaling behavior predicted
by Eq.~(\ref{qtscaN}).

The Loschmidt echo is expected to vanishes in the
large-time limit, due to the fact that the particles escape from the
trap in their free expansion. This is also formally obtained by noting
that the large-$t$ behavior of the one-particle wave functions,
cf. Eq.~(\ref{fexev}), have the following asymptotic behavior
\begin{equation}
\psi_n(x,t) \approx  \sqrt{2\over \pi^3 t} {1 - (-1)^n\over  n} \,,
\label{asybeh}
\end{equation}
i.e. they tend to be independent of $x$, corresponding to the fact
that when $x\ll v_F t$ (where $v_F$ is the Fermi velocity) the
one-particle wave functions within the 
space occupied initially can be approximated by
a constant.

Results for the Loschmidt echo are shown in Fig.~\ref{hdquench},
up to $N=200$ particles, for some values of $\tau$. They show that
in the large-$N$ limit the echo function increases as
\begin{equation}
Q(\ell_0,t,N) \sim N \ln N\,.
\label{largeNfreex}
\end{equation}
Therefore, in the case of hard-wall trap, the echo function $Q$ turns
out to increases more slowly than the case of harmonic traps,
cf. Eq.~(\ref{lochecho}).

\section{Interacting fermion gases}
\label{ifs}

\subsection{The Hubbard model}
\label{hubbard}

We now discuss the effects of particle interactions on the
particle-number scaling behaviors obtained for free fermions,
in particular for the quantum work statistics. For this
purpose, we consider the Hubbard model describing lattice gases of
spinful fermions.  The Hamiltonian of the Hubbard model reads
\begin{eqnarray}
H_{\rm h} =  - t \sum_{\sigma,\langle {\bm x}{\bm y}\rangle} 
(c_{\sigma {\bm x}}^\dagger c_{\sigma {\bm y}} + {\rm h.c.})
+ U \sum_{\bm x} n_{\uparrow{\bm x}} n_{\downarrow{\bm x}}\,,
\label{hm}
\end{eqnarray}
where ${\bm x}$ are the sites of a cubic lattice, $\langle {\bm x}{\bm
  y}\rangle$ indicates nearest-neighbor sites, $c_{\sigma{\bm x}}$ is
a fermionic operator, $\sigma=\uparrow\downarrow$ labels the spin
states, and $n_{\sigma{\bm x}}\equiv c_{\sigma{\bm x}}^\dagger
c_{\sigma{\bm x}}$.  Again we set $t=1$.  Analogously to
noninteracting lattice Fermi gases, cf. Eq.~(\ref{freef}), the
external force trapping the particles is taken into account by adding
a potential term, i.e,
\begin{equation}
H = H_{\rm h} + H_{\rm v}\,,\quad H_{\rm v} = \sum_{\sigma,{\bm x}} V({\bm
  x},\ell) \,n_{\sigma{\bm x}}\,.
\label{hedef}
\end{equation}
The particle number operators $\hat{N}_\sigma= \sum_{\bm x}
n_{\sigma{\bm x}}$ are conserved, i.e., $[H,\hat{N}_\sigma]=0$.  For
simplicity, in the following we consider balanced Fermi systems, thus
$N_\uparrow=N_\downarrow=N/2$ where $N$ is the total number of
particles.  In this symmetric case $\langle n_{\uparrow{\bm x}}\rangle
= \langle n_{\downarrow{\bm x}}\rangle$ and $\langle c_{\uparrow{\bm
    x}}^\dagger c_{\uparrow{\bm y}} \rangle = \langle
c_{\downarrow{\bm x}}^\dagger c_{\downarrow{\bm y}} \rangle$.

We again consider an out-of-equilibrium dynamics arising from the
sudden change of the trap size, from $\ell_0$ to $\ell_1>\ell_0$, in
the dilute regime.  Our purpose is to discuss the particle-number
dependence of the average work and its fluctuations, associated with
this process.

\subsection{The dilute regime of the Hubbard model}
\label{dilregime}

In order to investigate the particle-number dependence of the work
fluctuations, we need to summarize a number of known results
concerning the equilibrium correlation functions of $N$-particle
interacting lattice fermions associated with their ground state in the
presence of an external power-law potential trapping them, which show
a corresponding equilibrium trap-size scaling.

\subsubsection{Three-dimensional systems}
\label{3dhub}

In the language of the renormalization-group theory, the power-law
scaling behaviors in the dilute regime are controlled by a corresponding
dilute fixed point, related to the vacuum-to-metal
quantum transition~\cite{Sachdev-book}.
The renormalization-group analysis
of the effects of the interactions shows that the $U$ term is
irrelevant at the dilute fixed point for $d>2$, because its RG
dimension $y_U=2-d$ is negative. Therefore, the asymptotic trap-size
dependence in the dilute regime turns out to be the same as that of a
free Fermi gases of $N$ particles with $N_\uparrow=N_\downarrow=N/2$,
independently of $U$, at least for $U>U^*$ with
$U^*<0$~\cite{Sachdev-book}.  The corresponding trap-size scaling
reads~\cite{ACV-14}
\begin{eqnarray}
&&\rho({\bm x},\ell,U,N) \approx \ell^{-3\theta} \,2\, S({\bm X},N/2)\, ,
\label{rho3d}\\ 
&&C({\bm x}_1,{\bm x}_2,\ell,U,N)  \approx
\ell^{-3\theta} \, 2\, E({\bm X}_1,{\bm X}_2,N/2) \,,
\nonumber\\
&&G({\bm x}_1,{\bm x}_2,\ell,U,N)  \approx \ell^{-6\theta} 
\,2\, Z({\bm X}_1,{\bm X}_2,N/2)\,,
  \nonumber\\
  &&{\bm X}_i \equiv {\bm x}_i/\ell^\theta,\qquad \theta = {p\over p+2}\,,
  \nonumber
\end{eqnarray}
for the particle density, the one-particle and connected
density-density correlations, respectively.  The scaling functions
$S$, $E$ and $Z$ are the same trap-size scaling functions of the free
fermion theory.  The presence of the on-site interaction 
associated with the parameter $U$ induces
$O(\ell^{-(d-2)\theta})$ scaling corrections.
They dominate the scaling corrections expected within the lattice
model of free spinless fermion, i.e., the Hubbard model with $U=0$,
which are relatively suppressed as
$O(\ell^{-2\theta})$.~\cite{CV-10-BH}

\subsubsection{Lower-dimensional systems}
\label{21dhub}

The on-site on-site coupling $U$
becomes marginal in two dimensions, indeed its
renormalization-group dimension $y_U=2-d$ vanishes, thus a residual
weak dependence on $U$ is expected in the asymptotic regime. More
precisely we expect~\cite{ACV-14,Nigro-17}
\begin{eqnarray}
&&\rho({\bm x},\ell,U,N) \approx \ell^{-2\theta} {\cal R}({\bm X},U,N),
\label{rho2d}\\ 
&&C({\bm x}_1,{\bm x}_2,\ell,U,N)  \approx 
\ell^{-2\theta}  {\cal C}({\bm X}_1,{\bm X}_2,U,N)  \,,
\nonumber\\
&&G({\bm x}_1,{\bm x}_2,\ell,U,N)  \approx  \ell^{-4\theta} 
{\cal G}({\bm X}_1,{\bm X}_2,U,N)\,.\nonumber
\end{eqnarray}

Finally, in one dimension the $U$ term turns out to be relevant, since
$y_U=1$, therefore the asymptotic behaviors are expected to change.
The relevance of the $U$ term in one dimension gives rise to nontrivial
asymptotic trap-size scaling limits, requiring an appropriate
rescaling of the parameter $U$. This is taken into account 
by introducing the scaling variable
\begin{equation}
U_r = U \ell^{\theta}\,,
\label{urdef}
\end{equation}
where 
$\theta$ is the same exponent of Eq.~(\ref{thetaexp}).  
Indeed, the system develops the
trap-size scaling behavior
\begin{eqnarray}
&&\rho(x)  \approx \ell^{-\theta}  {\cal R}(X,U_r,N)\,, 
\label{rho1d}\\ 
&&C(x_1,x_2) \approx \ell^{-\theta}  {\cal C}(X_1,X_2,U_r,N)\, ,
\nonumber\\
&&G(x_1,x_2)\approx \ell^{-2\theta} 
{\cal G}(X_1,X_2,U_r,N)\,,
\nonumber 
\end{eqnarray}
where $X_i=x_i/\ell^{\theta}$.
These scaling behaviors are expected to be approached with
power-law suppressed corrections.  Of course, for $U_r=0$, i.e., for a
strictly vanishing $U$, we must recover the scaling functions of the
free Fermi gas, taking into account that an unpolarized free Fermi
gases of $N$ particles is equivalent to two independent spinless Fermi
gases of $N/2$ particles.

\subsubsection{The continuum limit}
\label{contlim}

It is important to note again that the trap-size scaling limit
corresponds to a continuum limit in the presence of the trap, i.e., it
generally realizes a continuum quantum field theory in the presence of
an inhomogeneous external field~\cite{CV-10,ACV-14}.  In particular,
in the trap-size scaling limit the observables of the one-dimensional
trapped Hubbard model can be written in terms of the solutions he
continuum Hamiltonian~\cite{ACV-14,Nigro-17},
\begin{eqnarray}
H_c = \sum_{i=1}^N \left[ {{\bm p}_i^2\over 2 m} + V({\bm x}_i)
  \right] + g \sum_{i=1}^{N_\uparrow} \sum_{j=1}^{N_\downarrow}
\delta({\bm x}_i-{\bm x}_j)\,,\quad
\label{contmodels}
\end{eqnarray}
describing $N$ fermions, with $N_\uparrow = N_\downarrow=N/2$,
interacting through a local $\delta$-like term.  In particular, in one
dimension we recover the so-called Gaudin-Yang
model~\cite{Gaudin-67,Yang-67}.  Indeed, the trap-size scaling limit
of the one-dimensional Hubbard model at fixed $N$ is related to the
Gaudin-Yang model with $g \sim U_r\equiv U\ell^\theta$.  More
precisely, the trap-size scaling functions entering formulas
(\ref{rho1d}) are exactly given by corresponding quantities of the
Gaudin-Yang problem with a trap of unit size.  Analogously in two
dimensions we recover the continuum interacting model with $g\sim
U$.~\cite{Nigro-17,footnoteregu} Finally, in three dimensions, the
continuum limit is given by the trap-size scaling of the free Fermi
theory, for any value of the lattice coupling $U>U^*$ with
$U^*<0$~\cite{Sachdev-book}.

\subsection{Particle-number dependence of the quantum work}
\label{pahumo}

We now discuss the particle-number dependence of the quantum work
associated with a sudden quench of the trap size, from $\ell_0$ to
$\ell_1>\ell_0$, starting the ground state of the Fermi gas in the
trap of size $\ell_0$. We again consider the definition of work
distribution given by Eq.~(\ref{pwdefft}), in particular
Eq.~(\ref{pwdef}). To compute the average work we follow the same line
of reasoning used in the case of free Fermi gases, see
Sec.~\ref{avwork}. This leads us to the following general formula for
the trap-size scaling limit of the average work,
\begin{eqnarray}
&&\langle W\rangle \approx \ell_0^{-2\theta}\, B(\delta_\ell)\,
  A_1(U_r,N)\,,
\label{workhub}
\\
&& A_1(U_r,N) = \int d{\bm X} |{\bm X}|^p {\cal R}({\bm X},U_r,N)\,,
\nonumber
\end{eqnarray}
where ${\bm X} = {\bm x}/\ell_0^\theta$, ${\cal R}({\bm X},U_r,
\ell_0)$ is the rescaled particle density of the ground state with
trap size $\ell_0$, and $B(\delta_\ell)$ is defined in
Eq.~(\ref{w1tss}).  Analogously for the square work fluctuations,
following the initial steps outlined in Sec.~\ref{workflu}, we obtain
\begin{eqnarray}
&&\langle W^2\rangle_c \approx 
\ell_0^{-4\theta} \,B(\delta_\ell)^2\, A_2(U_r,N)\,,
\label{work2hub}\\
&&A_2(U_r,N)=
\int d{\bm X}_1 d{\bm X}_2 |{\bm X}_1|^p |{\bm X}_2|^p
      {\cal G}({\bm X}_1,{\bm X}_2,U_r,N) \,, \nonumber
\end{eqnarray}
where ${\cal G}$ is the rescaled density-density connected correlation
function.  As shown in Sec.~\ref{dilregime}, the effective on-site
coupling $U_r$, is given by
\begin{eqnarray}
&U_r = U\ell_0^\theta \quad &{\rm for} \;\;d=1\,,\label{urd}\\
&U_r = U     \quad &{\rm for} \;\;d=2\,,\nonumber\\
&U_r = 0  \quad &{\rm for} \;\;d=3\,.\nonumber
\end{eqnarray}

We now argue that the power laws associated with the particle-number
dependence of the average work and its square fluctuations are 
generally analogous to
those of the $d$-dimensional free Fermi gases.

In the case of three-dimensional Fermi gases this claim is clearly
a consequence of the fact that in the trap-size scaling 
functions ${\cal R}$ and ${\cal G}$ coincide with those of the free
Fermi gases, cf. Eqs.~(\ref{rho3d}), when the onsite coupling $U$ is
larger than a negative number $U^*<0$.  Therefore
Eqs.~(\ref{largeNw1}), (\ref{tssw2}) and (\ref{Jp2dp2}) are expected
to hold as well.

On the other hand, as shown by Eqs.~(\ref{rho2d}) and (\ref{rho1d}),
the trap-size scaling, or continuum limit, of lower-dimensional models
is more complicated. Let us first discuss the apparently more
complicated case of one-dimensional systems, whose continuum limit
corresponds to the Gaudin-Yang model.  As shown in Ref.~\cite{ACV-14},
for a large number of particles (still remaining in the dilute
regime), the trap-size scaling function of the particle density
behaves asymptotically as
\begin{eqnarray}
{\cal R}(X,U_r,N) \approx N^{1/2} {\cal R}_\infty(X/N^{1/2},U_r/N^{1/2}) 
\label{lnrho1}
\end{eqnarray}
where ${\cal R}_\infty(z,u)$ is a nontrivial scaling function, and
power-law suppressed corrections are neglected.  This already suggests
that the effect of a finite continuum coupling $U_r$ gets suppressed
in the large-$N$ limit. As we shall see, this is also confirmed by
arguments based the relation between the trap-size scaling of the
trapped Hubbard model and the continuum Gaudin-Yang model, which
allows us to determine the trap-size scaling functions of the particle
density and its correlation, i.e., ${\cal R}(X,U_r,N)$ and ${\cal
  G}(X_1,X_2,U_r,N)$ respectively, in the strongly repulsive and
attractive limits, i.e., $U_r\to\infty$ and $U_r\to -\infty$.

The equation of state of the homogenous Gaudin-Yang model is exactly
known for both repulsive and attractive zero-range
interaction~\cite{Gaudin-67,Yang-67}.  It is characterized by
different asymptotic regimes with respect to the effective
dimensionless coupling $\gamma \equiv g/\rho$, where $\rho$ is the
particle density. At weak coupling $\gamma \ll 1 $ it behaves as a
perfect Fermi gas; in the strongly repulsive regime, $\gamma\gg 1$ the
equation of state approaches that of spinless Fermi gas; in the
strongly attractive regime $\gamma\to -\infty$ and for unpolarized
gases it matches that of a one-dimensional gas of impenetrable
bosons~\cite{Girardeau-60}, more precisely hard-core bosonic molecules
of fermion pairs~\cite{ABGP-04,FRZ-04}.  We know that in the
$g\to\infty$ limit the particle density and its correlations of the
Gaudin-Yang model become identical to those of a gas of $N$ spinless
fermions~\cite{Schulz-90,GCWM-09,GBL-13}. This would imply that the
$U_r\to \infty$ limit of the trap-size scaling functions is
\begin{eqnarray}
 &&{\cal R}(X,U_r\to\infty,N) = S(X,N), \label{rhoui}\\
 &&{\cal G}(X_1,X_2,U_r\to\infty,N) = Z(X_1,X_2,N), \quad
 \nonumber
\end{eqnarray}
where $S$ and $Z$ are the same functions entering the spinless
free-fermion trap-size scaling.

In the $g\to -\infty$ limit the density properties of the Gaudin-Yang
model is expected to match that of an ensemble of hard-core $N/2$
bosonic molecules constituted by up and down fermions.  Indeed, with
increasing attraction, the pairing becomes increasingly localized in
space, and eventually the paired fermions form a tightly bound bosonic
molecule.  Actually, the results of Ref.~\cite{ABGP-04} for harmonic
traps, see also Ref.~\cite{ACV-14}, 
show that these bound states get trapped in a smaller region,
with an effective trap size $\ell_b=\ell/2$ in the strongly attractive
limit.  Thus, we expect that in the $g\to -\infty$ limit the particle
density of the unpolarized Gaudin-Yang model with a harmonic trap
matches that of $N/2$ hard-core doubly-charged bosons with an
effective trap size $\ell_b= \ell/2$, which in turn can be mapped into
a free gas of $N/2$ spinless doubly-charged fermions in a harmonic
trap of size $\ell_b$.  On the basis of these arguments, the $U_r\to
-\infty$ limit of the trap-size scaling functions for harmonic traps
is expected to be
\begin{eqnarray}
 &&{\cal R}(X,U_r\to -\infty,N) = 2^{3/2} S(\sqrt{2} X,N/2), 
\label{rhouim}\\
 &&{\cal G}(X_1,X_2,U_r\to -\infty,N) = 8 Z(\sqrt{2} X_1,\sqrt{2}X_2,N/2).
\nonumber
\end{eqnarray}

These results for the Gaudin-Yang model imply that, if we compute the
average work (\ref{workhub}) and its fluctuations (\ref{work2hub}) in
the limits $U_r\to\pm\infty$, we obtain formulas analogous to those
for the free Fermi theory when inserting them into the corresponding
Eqs.~(\ref{rhoui}) and (\ref{rhouim}). In particular, we obtain the
same large-$N$ power laws, with trivial changes of their coefficients.
These arguments suggest that the large-$N$ behavior of one-dimensional
systems is essentially the same of the of free Fermi particles, at
least in the regime of trap-size scaling.

Another important issue concerns the degree of universality of the above
claims, with respect to further local interaction terms extending the
Hubbard model (\ref{hm}).
 This can be inferred by the universality of the behavior of
the particle density, and particle density correlations~\cite{ACV-14}.
We expect that they are universal with respect to a large class of
further short-ranged interaction terms, such as
\begin{equation}
H_{nn} = \sum_{\sigma,\sigma'} w_{\sigma\sigma'} 
\sum_{\langle{\bm x}{\bm y}\rangle}\, n_{\sigma {\bm x}}\, n_{\sigma'{\bm y}}.
\label{hddhm}
\end{equation}
Indeed, $H_{nn}$ may only give rise to a change of the effective
quartic coupling $U$ (when adding $H_{nn}$ to the Hubbard Hamiltonian,
the effective relevant quartic coupling becomes
$U+2w_{\uparrow\downarrow}$), and to further $O(l^{-\theta})$
corrections, due to the fact that they introduce other irrelevant RG
perturbations of renormalization-group dimension $y_w=-d$ at the
dilute fixed point.

In conclusion, the above arguments show that the large-$N$ power laws
of the work fluctuations remain unchanged when we consider
three-dimensional Fermi gases with short-ranged interactions with
positive on-site couplings (more precisely for $U>U^*$ with $U^*<0$).  We
also conjecture that this property extends to one-dimensional systems,
in the regime where trap-size scaling holds, and in particular in the
continuum Gaudin-Yang model for any interaction coupling.  We believe
that the same conclusion applies to two-dimensional systems for any
value of the on-site coupling $U$, for which the relation between the
trap-size scaling and continuum limit does not require a rescaling of
the coupling.

\section{Summary and conclusions}
\label{conclu}

We investigate the particle-number scaling behaviors characterizing
the out-of-equilibrium quantum dynamics of dilute $d$-dimensional
Fermi gases, in the limit of a large number $N$ of particles. We
consider protocols entailing variations of the external potential
constraining them within a limited spatial region, such as those
giving rise to a change of the size $\ell$ of the trap.  We consider
generic traps arising from external power-law potential, in particular
the case of harmonic traps and hard-wall traps.  We mostly consider
lattice gas models of noninteracting Fermi particles in the dilute
regime, $\ell/a\gg 1$ (where $a$ is the lattice spacing) and
$N/\ell^d\ll 1$, corresponding to the large trap-size limit keeping
$N$ fixed.  In the framework of the trap-size scaling, the asymptotic
large-$\ell$ behavior can be related to that of a continuum many-body
theory of Fermi particles in an external confining
potential~\cite{ACV-14,Nigro-17}.  Therefore, our results apply to
lattice Fermi gases in the dilute limit, and also to continuum Fermi
models such as the Gaudin-Yang model~\cite{Gaudin-67,Yang-67}.

We determine the asymptotic large-$N$ power laws of some features
characterizing the out-of-equilibrium dynamics of Fermi gases, arising
from the change of the trap features, starting from the equilibrium
ground state for the initial trap size $\ell_0$.  We focus on a number
of global quantities, providing information on the evolution of the
quantum state with respect to the initial one.  We consider the
ground-state fidelity associated with adiabatic changes of the trap
size, the quantum work average and its fluctuations associated with a
sudden change of the trap size, and the overlap of the quantum state
at a given time $t$ with the initial ground-state state as measured by
the so-called Loschmidt echo.  In the case of the quantum work
statistics, we also discuss the effects of short-ranged particle
interactions, in the framework of the Hubbard model and its continuum
limit realized in the trap-size scaling limit.

We show that the $N$ dependence of the first few moments of the work
statistics, associated with the sudden change of the trap size, can be
obtained from the scaling behaviors of the ground-state particle
density and its correlations, see Secs.~\ref{work} and \ref{ifs}.  Our
main results concern the asymptotic large-$N$ power laws for
$d$-dimensional Fermi gases in the dulute regime, confined by a
generic power-law potential.  The large-$N$ behavior of the average
work turns out to be
\begin{eqnarray}
\langle W \rangle \sim N^{1+2\theta/d}\,, \label{avwogen}
\end{eqnarray}
where $\theta=p/(p+2)$ and $p$ is the power law of the spatial
dependence of the confining potential, cf. Eq.~(\ref{potential}).
Analogous power laws are obtained for the square work fluctuations.
It is important to note
that the asymptotic large-$N$ behaviors that we consider should be
always intended within the dilute regime of the lattice gas models,
i.e., when the condition $N/\ell^d\ll 1$ is satisfied. The order of
the limits $\ell_0\to\infty$ and then $N\to\infty$ is essential, they
cannot be interchanged.

We also argue that short-ranged particle interactions, such as those
described by the Hubbard model and the Gaudin-Yang model, do not
change the large-$N$ power laws in the dilute regime, within
appropriate ranges of their coupling values, depending on the spatial
dimensions, see Sec.~\ref{pahumo}. In particular, for
three-dimensional systems the large-$N$ behavior is expected to be the
same of the free Fermi gases for on-site couplings $U$ larger than a
negative value $U^*<0$, thus including an interval around $U=0$ and
for any repulsive interaction. For one-dimensional models we argue
that the large-$N$ behaviors remain unchanged in the regime of
trap-size scaling, thus for the corresponding continuum Gaudin-Yang
model.

We note that, in the case of one-dimensional systems, the results for
non-interacting Fermi gases extends to one-dimensional Bose gases in
the limit of strong short-ranged repulsive interactions.  The basic
model to describe the many-body features of a boson gas confined to an
effective one-dimensional geometry is the Lieb-Liniger model with an
effective two-particle repulsive contact interaction~\cite{LL-63}.
The limit of infinitely strong repulsive interactions corresponds to a
one-dimensional gas of impenetrable bosons~\cite{Girardeau-60}, the
Tonks-Girardeau gas.  One-dimensional Bose gases with repulsive
two-particle short-ranged interactions become more and more nonideal
with decreasing the particle density, acquiring fermion-like
properties, so that the one-dimensional gas of impenetrable bosons is
expected to provide an effective description of the low-density regime
of confined one-dimensional bosonic gases~\cite{PSW-00}.  Due to the
mapping between one-dimensional gases of impenetrable bosons and
spinless fermions, the particle density of hard-core bosons, and its
correlations, are identical to those of free fermion gases. Therefore,
the results of this paper for the work statistics apply to
one-dimensional repulsively interacting Bose gases as well, subject to
analogous dynamic protocols.

For one-dimensional Fermi gases we also study the quantum evolution
arising from the change of the trap size, including the extreme case
of the free expansion of the gas after the drop of the trap. In the
case of harmonic traps, we present results for generic time
dependences of the trap size. We show that the particle-number
dependence of the echo function $Q=-\ln L_E$, where $L_E$ is the
Loschmidt echo, is generally characterized by the power-law behavior
\begin{equation}
Q = -\ln | \langle \Psi({\bm x}_1,...,{\bm
  x}_N;t) | \Psi({\bm x}_1,...,{\bm x}_N;t=0) \rangle|
\sim N^2\,,
\label{legen}
\end{equation}
independently of the particular protocol varying the trap.  This is
compared with the asymptotic behavior obtained when dropping a hard
wall, which turns out to increase more slowly, i.e. $Q\sim N\ln N$.

Quite remarkably, the particle-number scaling behaviors outlined in
this paper can be observed for systems with a relatively small number
of particles, i.e., $O(10^2)$ or even less.  Therefore, even systems
with relatively few particles may show definite signatures of the
scaling laws derived in this work. In this respect, present-day
quantum-simulation platforms have already demonstrated their
capability to reproduce and control the dynamics of ultracold atoms in
optical lattices, therefore the properties of the quantum many-body
physics discussed here may be tested with a minimal number of
controllable objects.  In particular, the work statistics may be
accessable experimentally in ultracold-atom systems, see, e.g.,
Refs.~\cite{HSDL-08,Dorner-etal-13,MDP-13}.

\appendix

\section{Ground state of Fermi gases}
\label{groundstate}

The ground state of a Fermi gas constituted by $N$ particles is given
by a Slater determinant, 
\begin{equation}
\Psi({\bm x}_1,...,{\bm x}_N) = {1\over \sqrt{N!}} {\rm det}
    [\psi_i({\bm x}_j)],
\label{fpsi2}
\end{equation}
where $\psi_i({\bm x})$ are the lowest $N$ eigensolutions of the
one-particle Schr\"odinger equation $H\psi_i=E_i \psi_i$.

In the case of the harmonic potential, the one-particle energy
spectrum in harmonic traps is discrete.  The eigensolutions can be
written as a product of eigenfunctions of corresponding
one-dimensional Sch\"rodinger problems, i.e.
\begin{eqnarray}
&&\psi_{n_1,n_2,...,n_d}({\bf x}) = \prod_{i=1}^d \phi_{n_i}(x_i),\quad
\label{prodfunc}\\
&&E_{n_1,n_2,...,n_d}= \sum_{i=1}^d e_{n_i},
\nonumber
\end{eqnarray}
where the subscript $n_i$ labels the eigenfunctions along the $d$
directions, which are
\begin{eqnarray}
&&\phi_n(x) = \xi^{-1/2}{H_{n-1}(X)\over \pi^{1/4} 2^{(n-1)/2}
    (n-1)!^{1/2}} \, e^{-X^2/2},\quad
 \label{1deigf}\\
&&\xi = \ell^{1/2},\qquad X = x/\xi\,,\nonumber \\ 
&&e_{n} = \ell^{-1} (n - 1/2), \quad
 n=1,2,... \nonumber
\end{eqnarray}
where $H_n(x)$ are the Hermite polynomials.  Note however that,
although the spatial dependence of the one-particle eigenfunctions is
decoupled along the various directions, fermion gases in different
dimensions present notable differences due to the nontrivial filling
of the lowest $N$ states which provides the ground state of the
$N$-particle system.  Exploiting the properties of the Hermite
polynomials, the ground state (\ref{fpsi}) of one-dimensional systems
with $N$ particles can be written as in Eq.~(\ref{phisolu}).

In the case of a hard-wall trap, corresponding to finite-volume
systems with open boundary conditions, the eigensolutions can be
written as a product of eigenfunctions of the corresponding
one-dimensional Schr\"odinger problem, analogously
to Eqs.~(\ref{prodfunc}) with  
\begin{eqnarray}
\phi_n(x) = \ell^{-1/2} {\rm sin}\left(n \pi 
{x+\ell\over 2\ell}\right),\quad
e_{n} =  \ell^{-2} \,{\pi^2\over 8} n^2, \quad\label{1deigfinf}
\end{eqnarray}
for $n=1,2,...$.

\end{document}